\documentclass[%
 reprint,
 amsmath,amssymb,
 aps,
pra,
]{revtex4-2}

\usepackage{graphicx}
\usepackage{dcolumn}
\usepackage{bm}
\usepackage{physics}
\usepackage{multirow}

\begin{document}


\title{Synergistic Dynamical Decoupling and Circuit Design for Enhanced Algorithm Performance on Near-Term Quantum Devices}

\author{Yanjun Ji}
\email{yanjun.ji@informatik.uni-stuttgart.de}
\affiliation{Institute of Computer Architecture and Computer Engineering, University of Stuttgart, Pfaffenwaldring 47, 70569 Stuttgart, Germany}

\author{Ilia Polian}
\email{ilia.polian@informatik.uni-stuttgart.de}
\affiliation{Institute of Computer Architecture and Computer Engineering, University of Stuttgart, Pfaffenwaldring 47, 70569 Stuttgart, Germany}

\date{\today}

\begin{abstract}

Dynamical decoupling (DD) is a promising technique for mitigating errors in near-term quantum devices. However, its effectiveness depends on both hardware characteristics and algorithm implementation details. This paper explores the synergistic effects of dynamical decoupling and optimized circuit design in maximizing the performance and robustness of algorithms on near-term quantum devices. By utilizing eight IBM quantum devices, we analyze how hardware features and algorithm design impact the effectiveness of DD for error mitigation. Our analysis takes into account factors such as circuit fidelity, scheduling duration, and hardware-native gate set. We also examine the influence of algorithmic implementation details, including specific gate decompositions, DD sequences, and optimization levels. The results reveal an inverse relationship between the effectiveness of DD and the inherent performance of the algorithm. Furthermore, we emphasize the importance of gate directionality and circuit symmetry in improving performance. This study offers valuable insights for optimizing DD protocols and circuit designs, highlighting the significance of a holistic approach that leverages both hardware features and algorithm design for the high-quality and reliable execution of near-term quantum algorithms.

\end{abstract}

\maketitle


\section{Introduction}

Near-term quantum (NISQ) devices \cite{preskill2018quantum} hold immense potential but face hurdles in accuracy and reliability due to inherent noise arising from environmental fluctuations, imperfect gate operations, and~qubit interactions. Moreover, limitations in qubit count and connectivity restrict the complexity of achievable quantum circuits. Robust error mitigation techniques~\cite{cai2023quantum} are therefore crucial for unlocking the full potential of NISQ devices.
Dynamical decoupling (DD) \cite{suter2016colloquium, ahmed2013robustness, pokharel2018demonstration, souza2012robust} stands out as a powerful approach for NISQ devices due to its simplicity and low resource overhead. It mitigates decoherence errors by applying a carefully designed sequence of control pulses during idle periods of the qubits. These pulses effectively suppress the unwanted interaction between qubits and their environment, protecting the desired quantum state. DD has been demonstrated in various quantum systems, including spins~\cite{de2010universal,du2009preserving,farfurnik2016improving,farfurnik2015optimizing,merkel2021dynamical,medford2012scaling}, superconducting qubits~\cite{pokharel2018demonstration,tripathi2022suppression,bylander2011noise}, and~trapped ions~\cite{biercuk2009experimental}. The~effectiveness of DD extends beyond decoherence suppression as~it can also mitigate crosstalk~\cite{tripathi2022suppression,evert2024syncopated,zhou2023quantum,shirizly2024dissipative,seif2024suppressing} and coherent errors~\cite{qiu2021suppressing}.

Numerous DD sequences have been developed, with~prominent examples including Carr--Purcell (CP) \cite{carr1954effects}, Carr--Purcell--Meiboom--Gill (CPMG) \cite{meiboom2004modified}, XY4~\cite{maudsley1986modified, alvarez2012iterative, viola1999dynamical, souza2012effects}, KDD~\cite{souza2011robust}, and~Uhrig dynamical decoupling (UDD) \cite{uhrig2007keeping}. However, the~effectiveness of different sequences varies significantly~\cite{ezzell2023dynamical}. Prior work has shown that the CPMG sequence outperforms the CP sequence~\cite{ahmed2013robustness}. Additionally, higher-order sequences generally outperform lower-order sequences~\cite{ahmed2013robustness, ezzell2023dynamical}, but~optimizing pulse intervals within CPMG and XY4 sequences can achieve comparable performance~\cite{ezzell2023dynamical}. In~the context of the quantum approximate optimization algorithm (QAOA) \cite{farhi2014quantum,blekos2024review,spieldenner2023mean, zhou2020quantum,harrigan2021quantum}, DD sequences like CP, CPMG, and~XY4 significantly enhance performance, while KDD shows worse performance on IBM quantum processing units (QPUs)~\cite{niu2022effects}. Combining DD sequences with pulse-level optimization in QAOAs can further improve performance~\cite{niu2022effects}.
While DD sequences mitigate errors, these additional control pulses can also introduce errors such as gate infidelities or crosstalk.
To address this trade-off, adaptive approaches have been explored~\cite{das2021adapt}, which estimate the potential benefit of DD for each qubit combination and selectively apply DD to the subset that offers the most benefit.  Furthermore, recent work has explored the use of empirical learning schemes for enhancing DD effectiveness on quantum processors~\cite{tong2024empirical}, leveraging data-driven techniques to optimize the implementation of DD sequences, leading to improved error suppression~capabilities.

{Despite significant research on DD sequences, the~interplay between their effectiveness and specific algorithm implementations on real hardware remains underexplored. This study bridges this critical gap by investigating how factors such as transpilation efficiency~\cite{ji2022calibration,gokhale2020optimized,leymann2020bitter,huang2023near, ji2023algorithm}, circuit structure~\cite{ji2023improving, ji2023optimizing}, and~gate decompositions~\cite{vartiainen2004efficient,ji2023optimizing} influence DD performance on IBM QPUs. By~analyzing how these implementation details interact with the chosen DD sequence, we uncover synergistic effects that enhance overall algorithm performance on hardware. This focus on codesigning hardware and software offers deeper insights than separate studies of DD or individual circuit optimization techniques. We focus on the CPMG sequence due to its established effectiveness, robustness~\cite{ahmed2013robustness}, and~good performance for QAOAs on IBM QPUs~\cite{niu2022effects}.}

{Our analysis of results from eight IBM QPUs demonstrates} an inverse relationship between the initial performance of the algorithm without DD and the effectiveness of DD. While DD generally enhances performance and robustness, circuits with inherently higher fidelity and shorter execution times benefit less from DD than those with lower initial performance. Moreover, factors such as hardware-native gates of QPUs, the chosen gate decomposition strategy, and~the optimization level can also influence the effectiveness of DD. Additionally, using gates with consistent directionality and maintaining circuit symmetry during design lead to improved performance. These findings emphasize the importance of a holistic approach that considers both hardware and software optimization for the successful execution of algorithms with DD on NISQ devices, providing valuable insights for optimizing DD protocols and designing more robust quantum~algorithms.

This paper is structured as follows. Section~\ref{sec:method} outlines the methodology, presenting the benchmark circuits and metrics used to evaluate algorithm performance. Hardware considerations and a proposed synergistic design approach combining both hardware and algorithmic factors are also described. Section~\ref{sec:resul_analy} then analyzes the results, exploring the impact of various hardware factors, including circuit fidelity, schedule duration, and~native gate sets, as~well as algorithmic factors, including different implementations, DD sequences, and~optimization levels. Finally, Section~\ref{sec:discu_conclu} discusses the key findings and~concludes.

\section{Methodology\label{sec:method}}

We experimentally investigate the impact of hardware and algorithmic factors on the effectiveness of dynamical decoupling in superconducting quantum processes.
Hardware factors, such as circuit fidelity, schedule duration, and~native gate sets, define the fundamental capabilities of a quantum device. Conversely, algorithmic factors encompass the design choices made during algorithm implementation (specific sequence of quantum operations), error suppression strategy (selection of a DD sequence), and~circuit optimization techniques. These algorithmic choices ultimately determine how efficiently the inherent capabilities of the hardware are utilized for optimal performance. Understanding these factors can potentially improve the practical implementation of algorithms on near-term quantum~devices.

\subsection{Benchmark Circuits and~Metrics}

We first present benchmark circuits used in our demonstration and metrics employed to assess algorithm performance and DD~effectiveness.

\subsubsection{QAOA for Portfolio~Optimization}

We employ QAOAs for portfolio optimization as benchmarks.
The portfolio optimization problem aims to select the optimal portfolio from all available options to maximize expected returns while minimizing financial risk. The~QAOA has the potential to address this challenge~\cite{baker2022wasserstein, brandhofer2022benchmarking, egger2020quantum}. The~cost Hamiltonian, which describes the portfolio optimization problem for $n$ available assets, is expressed as~\cite{brandhofer2022benchmarking}
\begin{equation}
	\hat{H}_c= \sum_{i=1}^{n-1}\sum_{j=i+1}^n c_{i,j}\hat{Z}_i\hat{Z}_j-\sum_{i=1}^n k_i \hat{Z}_i.
	\label{eq:hamilt_cost}
\end{equation}
{The parameters $c_{i,j}$ and $k_i$ in Equation~(\ref{eq:hamilt_cost}) depend on factors specific to the portfolio optimization problem, such as the covariance matrix and the return vector. In~particular, $c_{i,j}=\frac{\lambda}{2}(q\sigma_{ij}+A)$ and $k_i=\frac{\lambda}{2} \left[A (2B-n)+(1-q)\mu_i-q \sum_{j=1}^n \sigma_{ij}\right]$, where $\lambda$ is the global scaling factor, $q$ is risk preference, $\sigma_{ij}$ is the covariance between assets $i$ and $j$, $A$ is the penalty factor, $B$ is the number of assets to be chosen, and~$\mu_i$ is the expected return of asset $i$. The~terms $\hat{Z}_i \hat{Z}_j$ and $\hat{Z}_i$ correspond to} the $ZZ$ interaction on qubits ($i$, $j$) and Pauli $Z$ operator acting on qubit $i$, respectively.
The mixer Hamiltonian is given by~\cite{farhi2014quantum}
\begin{equation}
	\hat{H}_m= \sum_{i=1}^{n}\hat{X}_i,
	\label{eq:hamilt_mixer}
\end{equation}
where $\hat{X}_i$ is the Pauli $X$ operator acting on qubit $i$.
After a QAOA depth of $p$, the~total system evolves to $\ket{\psi}=\prod_{j=1}^{p} e^{-i\gamma_j \hat{H}_c} e^{-i\beta_j \hat{H}_m} \ket{\psi_0}$, where $\ket{\psi_0}$ is the eigenstate of the mixer Hamiltonian. The~aim of the QAOA is to find $2p$ parameters ($\beta_1, ..., \beta_p, \gamma_1, ..., \gamma_p$) that minimize the expectation value of the cost Hamiltonian $F = \bra{\psi}\hat{H}_c\ket{\psi}$. We define the approximation ratio of the QAOA as
\begin{equation}
	r = \frac{F-F_{\max}}{F_0-F_{\max}},
	\label{eq:appr_ratio_portopt}
\end{equation}
where $F_0$ represents the optimal value and~$F_{\max}$ signifies the worst-case~value.

Our study uses a QAOA with qubit numbers ranging from 3 to 12 and a depth of 1. To~establish a baseline for experimental results, we present the simulation results conducted in a noise-free environment with Qiskit's Qasm simulator in Table~\ref{tab:qaoa_noise_free}. The version of Qiskit used throughout this paper is 0.45.3. All data use 30,000 circuit repetitions (shots). We observe that the approximation ratio and success probability decrease as the number of qubits~increases.

\begin{table*}[tb]
\caption{Noise-free simulation results of the quantum approximate optimization algorithm (QAOA) for portfolio optimization using Qiskit's Qasm simulator with 30,000~shots.\label{tab:qaoa_noise_free}}
\centering
    \begin{ruledtabular}

\begin{tabular}{lllllllllll}
{\textbf{Number of Qubits}}	& \textbf{3}	& \textbf{4}	& \textbf{5}	& \textbf{6}	& \textbf{7}	& \textbf{8}	& \textbf{9}	& \textbf{10}	& \textbf{11}	& \textbf{12}\\
\hline
{Approximation ratio} & 0.9751 & 0.4342 & 0.3776 & 0.3734 & 0.3589 & 0.3144 & 0.2806 & 0.3241 & 0.2933 & 0.3161\\
{Success probability} & 0.9747 & 0.1536 & 0.1131 & 0.0422 & 0.0227 & 0.0124 & 0.0065 & 0.0057 & 0.0018 & 0.0006\\
\end{tabular}
    \end{ruledtabular}

\end{table*}

\subsubsection{Metric~Definition\label{subsubsec:metri}}

To quantify the impact of noise, we introduce two normalized metrics: the normalized approximation ratio (NAR) and the normalized success probability (NSP). These metrics are defined as
\begin{eqnarray}
    \text{NAR} &=& r_\epsilon/r_0,\\
    \text{NSP} &=& p_\epsilon/p_0,
    \label{eq:NAR_NSP}
\end{eqnarray}
where $r_\epsilon$ and $p_\epsilon$ represent the approximation ratio and success probability obtained under noise conditions. $r_0$ and $p_0$ denote corresponding values obtained from the simulated noise-free case (Table~\ref{tab:qaoa_noise_free}). For~the same problem instances, we use identical values of $r_0$ and $p_0$ for evaluation. Due to noise, the NAR and NSP typically exhibit values less than one. However, in~rare instances, these metrics may exceed unity, indicating that specific noise patterns or randomness enhance performance compared to the simulated noise-free~case.

Without any error mitigation being applied, the NAR and NSP for algorithms executing on real noisy quantum devices are given by
\begin{eqnarray}
    \mathrm{NAR}_\mathrm{\scriptscriptstyle B} &=& r_b/r_0,\\
    \mathrm{NSP}_\mathrm{\scriptscriptstyle B} &=& p_b/p_0,
    \label{eq:NAR_NSP}
\end{eqnarray}
where B represents results obtained from real hardware without error mitigation. $r_b$ and $p_b$ represent the corresponding approximation ratio and success probability, respectively.
When a specific error mitigation strategy, such as the DD sequence, is applied, the NAR and NSP are given by
\begin{eqnarray}
    \mathrm{NAR}_\mathrm{\scriptscriptstyle DD} &=& r_{d}/r_0,\\
    \mathrm{NSP}_\mathrm{\scriptscriptstyle DD} &=& p_{d}/p_0,
    \label{eq:NAR_NSP_D}
\end{eqnarray}
where DD denotes the application of DD sequences. $r_d$ and $p_d$ represent the approximation ratio and success probability, respectively, with~error mitigation.
To assess DD effectiveness in error mitigation, we introduce two additional metrics: $\Delta_\mathrm{\scriptscriptstyle NAR}$ and $\Delta_\mathrm{\scriptscriptstyle NSP}$. These metrics are defined as the difference between the corresponding values obtained with and without DD~sequences:
\begin{eqnarray}
    \Delta_\mathrm{\scriptscriptstyle NAR} &= &\mathrm{NAR}_\mathrm{\scriptscriptstyle DD} - \mathrm{NAR}_\mathrm{\scriptscriptstyle B},\\
    \Delta_\mathrm{\scriptscriptstyle NSP} &= &\mathrm{NSP}_\mathrm{\scriptscriptstyle DD} - \mathrm{NSP}_\mathrm{\scriptscriptstyle B}.
    \label{eq:EDD}
\end{eqnarray}
A positive value for $\Delta_\mathrm{\scriptscriptstyle NAR}$ and $\Delta_\mathrm{\scriptscriptstyle NSP}$ indicates a successful improvement in performance resulting from the utilization of DD~sequences.

We further introduce the concept of error mitigation success rate (EMSR) to {quantify} the robustness of an error mitigation strategy. EMSR is defined as the percentage of experimental trials {where error mitigation improved the outcome compared to no mitigation}.
{A high EMSR indicates consistent performance improvement while a low EMSR suggests limited effectiveness or potential performance degradation. We employ two EMSR metrics in this study, $\mathrm{EMSR}_\mathrm{\scriptscriptstyle AR}$ and $\mathrm{EMSR}_\mathrm{\scriptscriptstyle SP}$, depending on whether the approximation ratio or success probability is used. Positive values of $\Delta_\mathrm{\scriptscriptstyle NAR}$ and $\Delta_\mathrm{\scriptscriptstyle NSP}$ contribute to increased $\mathrm{EMSR}_\mathrm{\scriptscriptstyle AR}$ and $\mathrm{EMSR}_\mathrm{\scriptscriptstyle SP}$, respectively.
}

\subsection{Hardware~Considerations\label{subsec:hardw_consid}}

This section provides information about the IBM quantum devices used in our experiments. The~QPUs with 27 qubits, namely ibmq\_mumbai, ibmq\_kolkata, ibm\_cairo, and~ibmq\_ehningen, operate using basis gates \{CX, ID, RZ, SX, X\}, where ID represents identity gate, RZ performs a single qubit rotation around the $z$-axis, X is the NOT gate, and~SX is the square root of X. On~the other hand, the~QPUs with 127 qubits, specifically ibm\_kyoto, ibm\_cusco, ibm\_brisbane, and~ibm\_sherbrooke, utilize basis gates \{ECR, ID, RZ, SX, X\}, where ECR is the echoed cross-resonance~gate.

\begin{figure}[tb]
\includegraphics[width=\columnwidth]{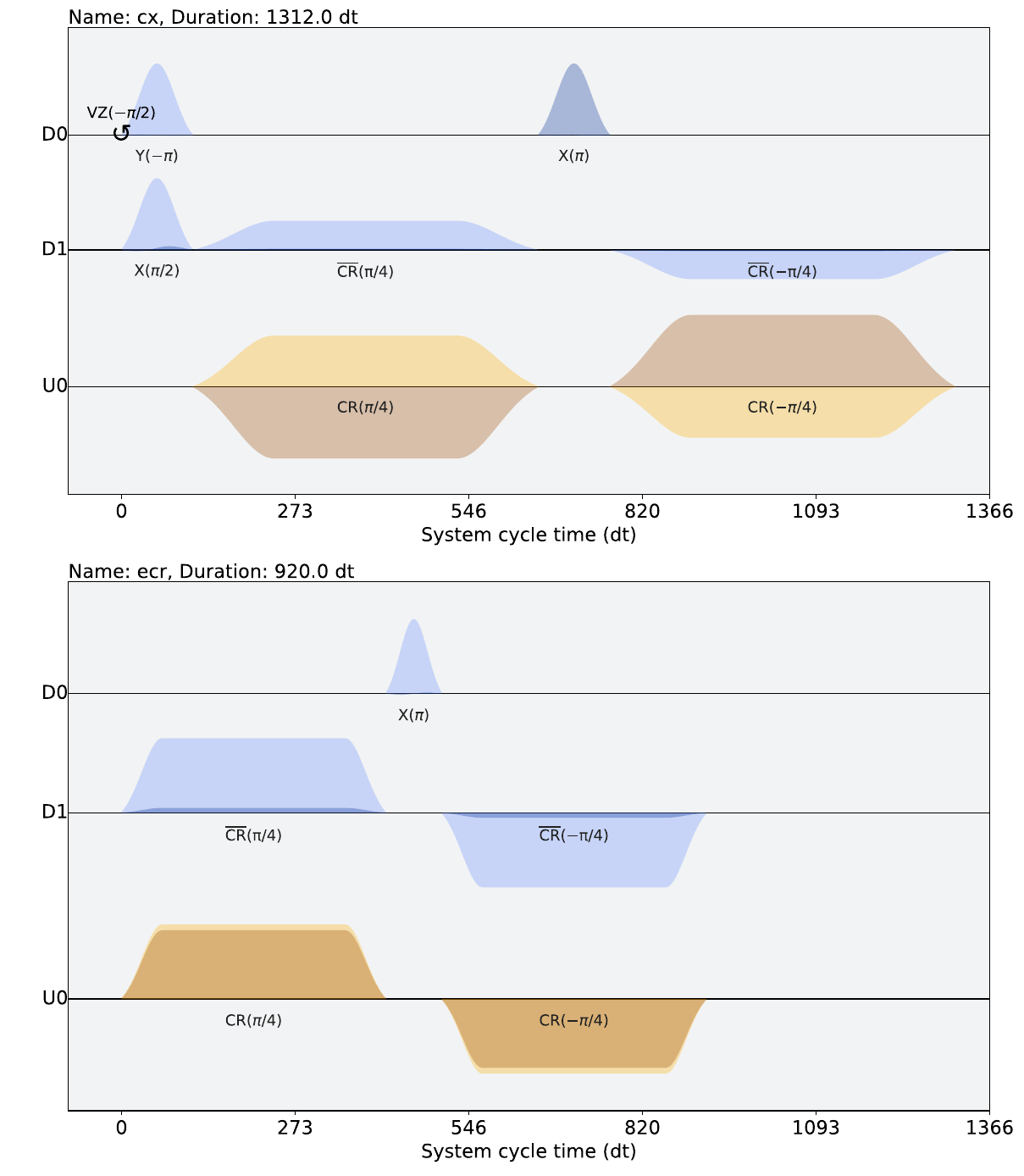}
\put(-\columnwidth,1.12\linewidth){(\textbf{a})}
\put(-\columnwidth,0.56\linewidth){(\textbf{b})}
\caption{\label{fig:ecr_cx_gate} Schedule of hardware-native two-qubit gates: (\textbf{a}) CX gate on qubit pair (0,1) of ibm\_cairo, and~(\textbf{b}) echoed cross-resonance (ECR) gate on qubit pair (0,1) of ibm\_cusco. The~system cycle time ($1 \mathrm{dt}$) is 2/9~ns $\approx$ 0.22~ns in ibm\_cairo, while it is 0.50~ns in ibm\_cusco. $D_i$ represents the drive channel acting on qubit $i$, and~$U_j$ is the control channel for a corresponding qubit pair $(c, t)$ driving the control qubit $c$ at the frequency of the target qubit $t$.}
\end{figure}

Figure~\ref{fig:ecr_cx_gate}(a) illustrates the schedule of a native CX gate on ibm\_cairo. This implementation employs a single-pulse gate duration of 112 dt and a cross-resonance (CR) gate duration of 544 dt, leading to a total duration of 1312 dt, where dt represents the system cycle time. In~contrast, Fig.~\ref{fig:ecr_cx_gate}(b) depicts the schedule of a native ECR gate on ibm\_cusco. Here, the~single-pulse and CR gate durations are 88 dt and 416 dt, respectively, resulting in a shorter total duration of 920 dt.
The CX gate is typically implemented using one ECR gate and multiple single-qubit gates.
A further decomposition of the CX gate into ECR and single-pulse gates at pulse level enables the elimination of single-pulse gates during circuit optimization~\cite{ji2023optimizing}.
Additionally, CX-based IBM QPUs, where CX is the native gate, support the operation of CX in two directions. On~the other hand, ECR-based QPUs, where ECR is the native gate, typically only support the ECR gate in one direction. Consequently, quantum circuits designed for these latter devices need to be decomposed into sequences of gates that include only the supported direction of the~ECR.

\vspace{-6pt}

\subsection{Synergistic Design~Approach\label{subsec:syner_design_appro}}

This section describes a synergistic design approach for maximizing the performance and robustness of algorithms on near-term quantum devices.
This approach acknowledges the critical interplay between the hardware's capabilities and the design choices made in the software implementation.
The quality of algorithm implementation directly affects the performance. Key aspects include the efficiency of transpilation processes, specific gate types used, and~the overall symmetry of the algorithm structure.
For instance, studies have shown that the algorithm-oriented qubit mapping (AOQMAP) method~\cite{ji2023algorithm} offers advantages in transpilation for variational quantum algorithms (VQAs) \cite{cerezo2021variational} compared to popular compilers such as Qiskit~\cite{qiskit} and Tket~\cite{sivarajah2020tket} by introducing fewer two-qubit gates, maintaining a shallower circuit depth, and~promoting higher symmetry~\cite{ji2023improving}.

In this study, we utilize the AOQMAP method~\cite{ji2023algorithm} to efficiently map circuits onto hardware, aiming to minimize SWAP gates and circuit depth on linear topologies.
Subsequently, we examine two implementations of QAOAs for portfolio optimization on CX-based IBM QPUs. These implementations differ in their choice of gate decompositions within the algorithms.
The first implementation, referred to as CX implementation, directly decomposes the gates in the QAOA into basis gates of the QPUs using Qiskit's transpiler with optimization level 3.
In comparison, the~second implementation, referred to as CZ implementation, initially decomposes gates in algorithms into CZ and single-qubit gates. Then, Qiskit's transpiler with optimization level 3 is used to perform optimization and decomposition into basis gates of QPUs. Previous studies have demonstrated that this CZ decomposition approach outperforms CX decomposition for ZZ and ZZ-SWAP gates in the QAOA on IBM QPUs~\cite{ji2023optimizing}.
During implementation, we also explore different optimization level settings in Qiskit and investigate their impact.
To ensure consistency in evaluations, we employ identical benchmark circuits and parameters. Additionally, we consistently use 30,000 shots for each demonstration. Within~the Qiskit framework, we default to using optimization level~3.

\begin{figure}[tb]
\includegraphics[width=\columnwidth]{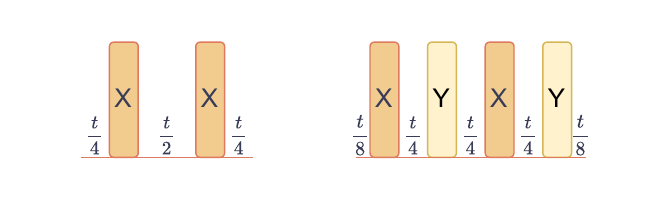}
\put(-0.95\columnwidth,0.22\columnwidth){(\textbf{a})}
\put(-0.56\columnwidth,0.22\columnwidth){(\textbf{b})}
\caption{\label{fig:dd_strategies} Two types of dynamical decoupling (DD) sequences: (\textbf{a}) Carr--Purcell--Meiboom--Gill (CPMG) and (\textbf{b}) XY4. The~delay time $t$ represents the idle time of the qubit minus the duration of the corresponding X or Y~pulses.}
\end{figure}

\begin{figure*}[tb]
 \includegraphics[width=0.97\linewidth]{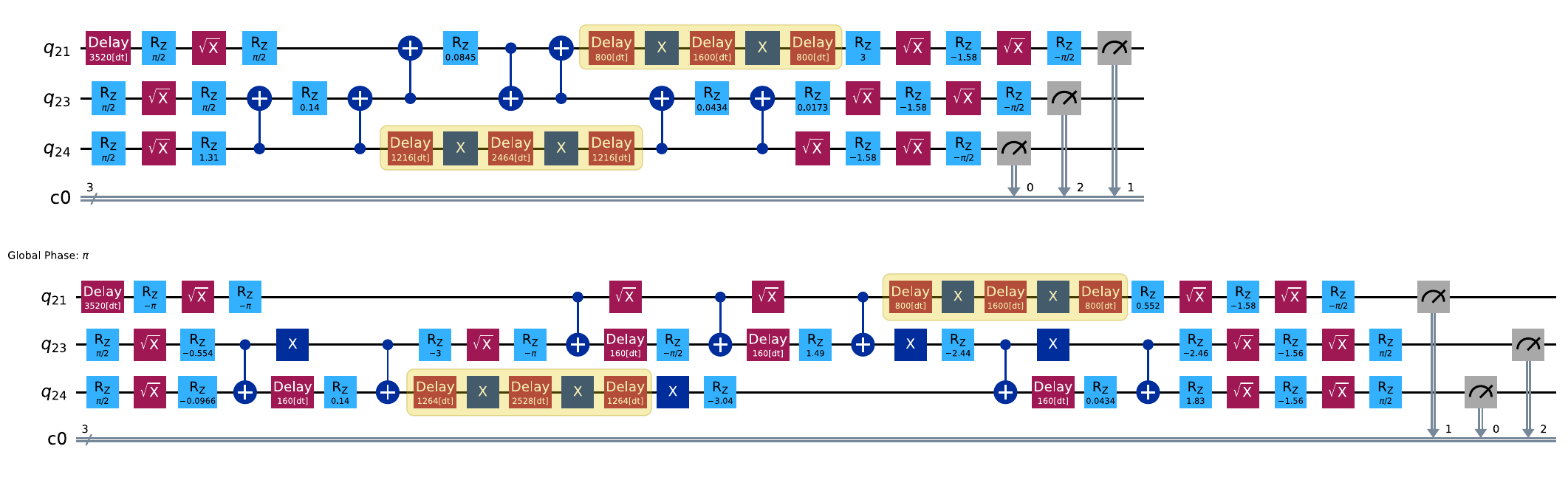}
    \put(-\linewidth,0.265\linewidth){{(\textbf{a})}}
    \put(-\linewidth,0.128\linewidth){{(\textbf{b})}}
    \caption{
    Two implementations of a three-qubit QAOA: (\textbf{a}) CX implementation and (\textbf{b}) CZ implementation, where both are decomposed and optimized using Qiskit’s transpiler with optimization level 3. Highlighted yellow boxes represent CPMG sequences.
    }
    \label{fig:circuit_algorithm_implementations}
\end{figure*}

We also investigate the effectiveness of two well-established DD sequences: CPMG and XY4.
As illustrated in Fig.~\ref{fig:dd_strategies}, the~CPMG sequence applies two X pulses separated by a delay of $\frac{t}{2}$, with~additional delays of $\frac{t}{4}$ at the beginning and end. The~parameter $t$ represents the time interval during which the qubit remains idle, excluding the duration of single-qubit pulses and, 
in~the case of CPMG, two X pulses. In~comparison, the~XY4 sequence utilizes two X and two Y pulses, each separated by a delay of $\frac{t}{4}$, with~additional delays of $\frac{t}{8}$ at the beginning and end.
Additionally, the~``alap'' (as late as possible) scheduling method, which schedules the stop time of instructions as late as possible, is used for scheduling gates and inserting DD sequences throughout our study.
Figure~\ref{fig:circuit_algorithm_implementations} showcases the resulting implementations of a three-qubit QAOA with a CPMG sequence on a 27-qubit QPU ibm\_kolkata using both CX and CZ implementations. Compared to CX implementation, CZ implementation employs all the same directed CX gates. Additionally, we observe an X gate inserted between control qubits of CX gates in the CZ implementation, potentially suppressing idle errors and improving performance. Moreover, CZ implementation shows improved symmetry compared to CX implementation.
By simultaneously optimizing both hardware and software aspects through careful algorithm design, efficient transpilation techniques, and~DD sequences, it is possible to fully exploit the capabilities of near-term quantum~algorithms.
\vspace{-12pt}

\vspace{-6pt}

\vspace{-6pt}

\section{Results~Analysis\label{sec:resul_analy}}

This section delves into the examination of multiple factors that influence algorithm performance and DD effectiveness. These factors are classified into two main categories: hardware and algorithmic.
More specifically, we thoroughly analyze the influence of circuit fidelity, schedule duration, and~native gate set on the performance of DD sequences. Furthermore, we explore the impact of algorithm implementation, choice of DD sequence, and~level of optimization on algorithm performance and DD effectiveness. The~objective of these analyses is to provide invaluable insights into the design and optimization of algorithm implementation for achieving efficient execution on quantum~devices.

\subsection{Impact of Hardware~Factors}

Our investigation begins by examining how hardware characteristics affect algorithm performance and DD effectiveness. The~CPMG sequence is chosen for our study due to its widespread adoption and relative simplicity, allowing us to gain fundamental insights. We analyze these factors using extensive datasets. The~experiments involve varying the qubit counts from 3 to 12 and investigating different combinations of algorithm implementations, including the CX and CZ versions of the QAOA, as~well as optimization levels 1 and 3 within the Qiskit framework. We conduct these experiments using eight QPUs, which consist of four 27-qubit devices---ibmq\_mumbai, ibmq\_kolkata, ibm\_cairo, and~ibmq\_ehningen---and~four 127-qubit devices: ibm\_kyoto, ibm\_cusco, ibm\_brisbane, and~ibm\_sherbrooke.
These extensive datasets provide a solid foundation for analyzing the impact of hardware factors on algorithm~performance.

\subsubsection{Circuit~Fidelity}

We first explore the impact of circuit fidelity on the performance. The~fidelity of a circuit $qc$, denoted as $\mathcal{F}_{qc}$, measures the agreement between the actual operation of a quantum circuit and its ideal operation. The~circuit fidelity can be mathematically represented as
\begin{equation}
\mathcal{F}_{qc} = \prod_{{G_s}\in qc} f_{{G_s}} \prod_{{G_t}\in qc} f_{{G_t}} \prod_{{G_m}\in qc} f_{{G_m}},
\end{equation}
where $f_{G_s}$, $f_{G_t}$, and~$f_{G_m}$ denote fidelities of a single-qubit gate $G_s$, a~two-qubit gate $G_t$, and~the measurement $G_m$, respectively, in~the circuit.
We focus on circuits with fidelities {between 0.01 and 1. This broader range of fidelities establishes a solid basis for investigating the effectiveness of DD sequences under realistic noise conditions. We analyze various metrics defined in Section~\ref{subsubsec:metri},} including $\mathrm{NAR}_\mathrm{\scriptscriptstyle B}$, $\mathrm{NAR}_\mathrm{\scriptscriptstyle DD}$, $\mathrm{NSP}_\mathrm{\scriptscriptstyle B}$, $\mathrm{NSP}_\mathrm{\scriptscriptstyle DD}$, $\Delta_\mathrm{\scriptscriptstyle NAR}$, and~$\Delta_\mathrm{\scriptscriptstyle NSP}$. The~measured data and the corresponding circuit fidelity are fitted using a linear function. The~correlation coefficient $C_r$ and $p$-value are computed to assess the quality of linear approximation. $C_r$ {measures the strength and direction of the linear relationship, ranging from $-1$ (perfect negative correlation) to 1 (perfect positive correlation), with~0 indicating no association.} The absolute value of $C_r$ {reflects the correlation strength: very strong (0.9--1.0), strong (0.7--0.9), moderate (0.4--0.7), weak (0.2--0.4), and~very weak (0--0.2).}
{I}t is important to note that correlation does not imply causation. The~$p$-value is a complementary statistical measure that evaluates the strength of evidence against the null hypothesis of no correlation. A~low $p$-value (typically below 0.05) suggests a statistically significant correlation, {possibly not due to random chance}. Linear fitting {builds on correlation by determining the best-fit line equation, enabling predictions based on the observed relationship.}

\begin{figure*}[tb]
\includegraphics[width=0.48\linewidth]{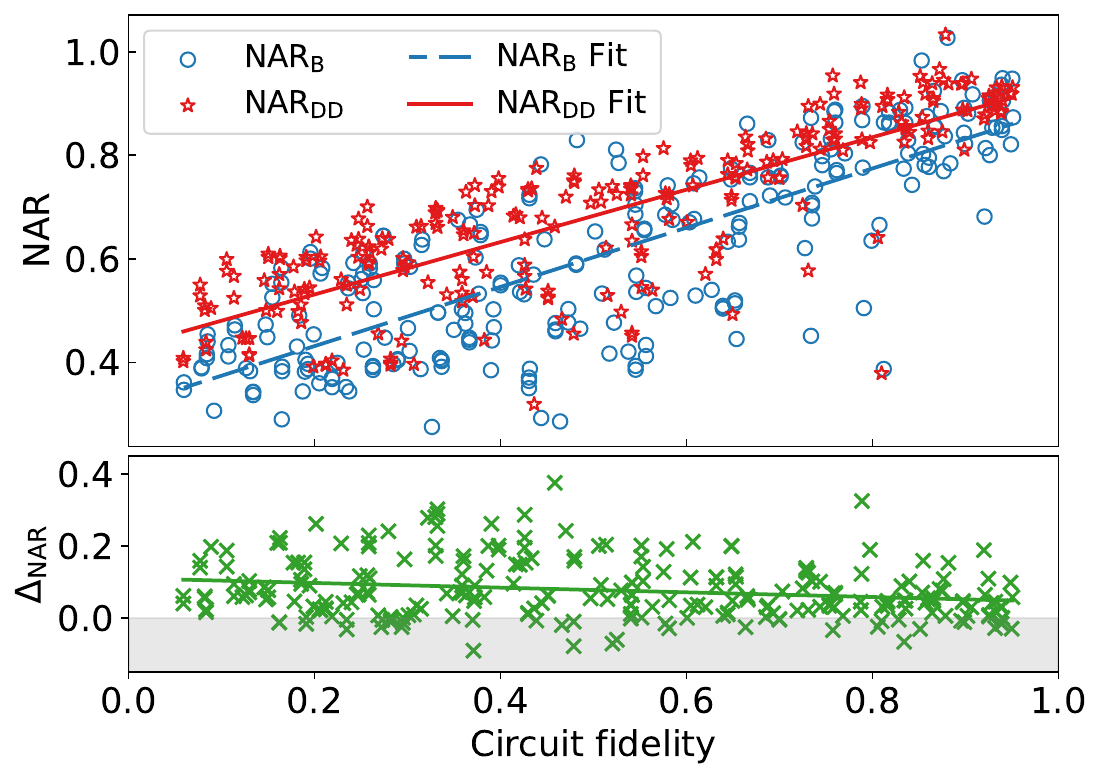}
\put(-0.48\linewidth,0.33\linewidth){(\textbf{a})}
\put(-0.48\linewidth,0.14\linewidth){(\textbf{c})}
\includegraphics[width=0.48\linewidth]{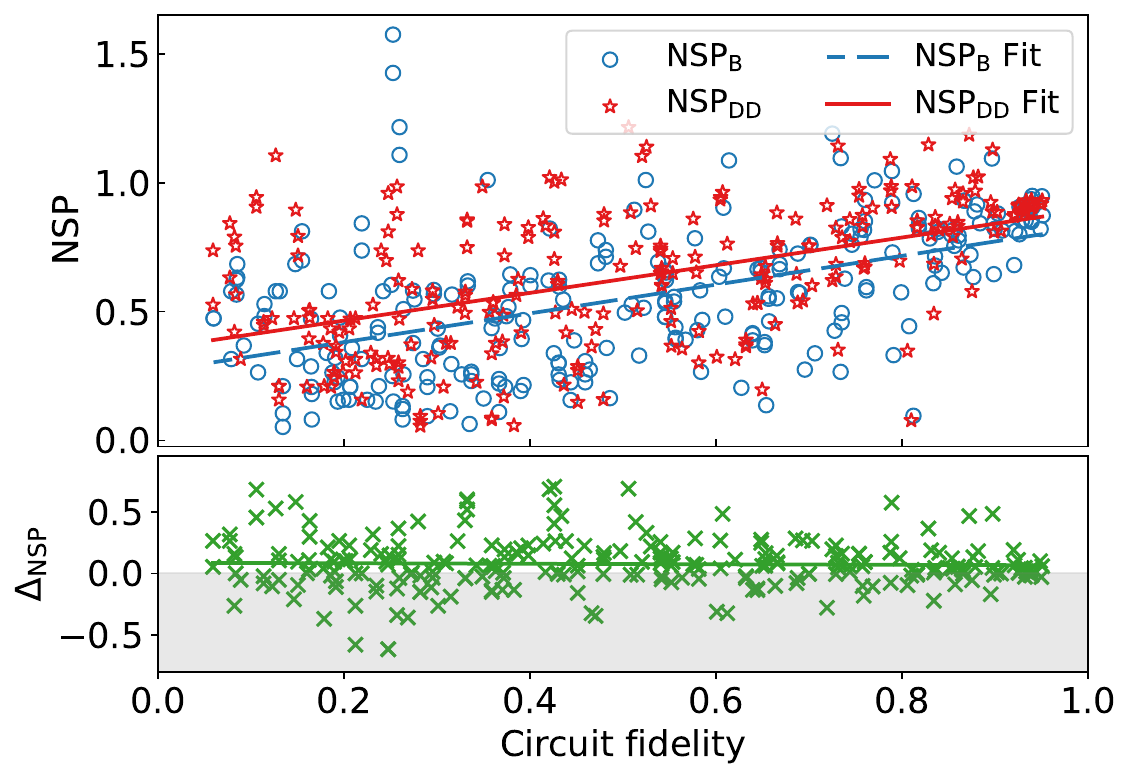}
\put(-0.48\linewidth,0.33\linewidth){(\textbf{b})}
\put(-0.48\linewidth,0.14\linewidth){(\textbf{d})}
\caption{\label{fig:circuit_fidelity}
Impact of circuit fidelity on algorithm performance and DD effectiveness: (a) normalized approximation ratio (NAR), (b) normalized success probability (NSP), (c) improvement in NAR after applying DD ($\Delta_\mathrm{\scriptscriptstyle NAR}$), and (d) improvement in NSP after applying DD ($\Delta_\mathrm{\scriptscriptstyle NSP}$).
Higher values of $\mathrm{NAR}_\mathrm{\scriptscriptstyle B}$, $\mathrm{NAR}_\mathrm{\scriptscriptstyle DD}$, $\mathrm{NSP}_\mathrm{\scriptscriptstyle B}$, and~$\mathrm{NSP}_\mathrm{\scriptscriptstyle DD}$ indicate better performance on actual quantum devices.
Values exceeding unity indicate that the performance achieved on the IBM quantum hardware surpasses the results obtained from the noise-free simulation. 
Positive values of $\Delta_\mathrm{\scriptscriptstyle NAR}$ and $\Delta_\mathrm{\scriptscriptstyle NSP}$ demonstrate improvements due to DD. The~CPMG sequence is used for all data points.
Each line in the graph represents a linear fit of the data.
The reported $\mathrm{EMSR}_\mathrm{\scriptscriptstyle AR}$ and $\mathrm{EMSR}_\mathrm{\scriptscriptstyle SP}$ are 85.55\% and 66.8\%, respectively.
}
\end{figure*}

\begin{table*}[tb]
    \caption{\label{tab:circ_fid_para} Parameters derived from the analysis of Fig.~\ref{fig:circuit_fidelity}.}
    \centering
    \begin{ruledtabular}
    \begin{tabular}{lllll}
     \textbf{Metric} & \textbf{Mean} & \textbf{Fit Function} & \textbf{Correlation Coefficient} & \textbf{\emph{p}-Value} \\
     \hline
     $\mathrm{NAR}_\mathrm{\scriptscriptstyle B}$ & 0.610 & $y = \phantom{-}0.572 x + 0.317$ &  $\phantom{-}0.809$ &  0 \\
     $\mathrm{NAR}_\mathrm{\scriptscriptstyle DD}$ & 0.687 & $y= \phantom{-}0.506x+0.430$ & $\phantom{-}0.827$  & 0 \\
     $\mathrm{NSP}_\mathrm{\scriptscriptstyle B}$ & 0.556 & $y= \phantom{-}0.558x+0.270$ & $\phantom{-}0.530$  &  0 \\
     $\mathrm{NSP}_\mathrm{\scriptscriptstyle DD}$ & 0.631 & $y= \phantom{-}0.537x+0.358$ & $\phantom{-}0.521$  &  0 \\
     \hline
     $\Delta_\mathrm{\scriptscriptstyle NAR}$ & 0.077 & $y=-0.065x+0.111$ &  $-0.209$ &  0.00075  \\
     $\Delta_\mathrm{\scriptscriptstyle NSP}$ & 0.075 & $y=-0.021x+0.086$ &  $-0.027$ &   0.66997\\
\end{tabular}
\end{ruledtabular}
\end{table*}

Figure~\ref{fig:circuit_fidelity} depicts a general trend of improved algorithm performance with increasing circuit fidelity. DD sequences enhance both the NAR and NSP on average. However, the NSP exhibits a wider range of variation for a given circuit fidelity compared to the approximation ratio, particularly at lower fidelities.
Table~\ref{tab:circ_fid_para} presents the average value, fitted function, correlation coefficient, and~$p$-value for each metric. A~stronger correlation is observed between NAR and circuit fidelity compared to NSP, suggesting a more pronounced dependence of NAR on fidelity. In~contrast, the~correlation between $\Delta_\mathrm{\scriptscriptstyle NSP}$ and circuit fidelity is very weak, as~evidenced by the low value of $C_r$ and high $p$-value. This suggests that the observed decrease in DD effectiveness might be due to random fluctuations or other factors not captured by circuit fidelity alone.
The average improvement in NAR and NSP due to DD sequences is approximately 0.08. Moreover, the~negative coefficients of the fitted lines for $\Delta_\mathrm{\scriptscriptstyle NAR}$ and $\Delta_\mathrm{\scriptscriptstyle NSP}$ suggest a potential decrease in DD effectiveness as circuit fidelity~increases.

\subsubsection{Schedule~Duration}

We now investigate the influence of schedule duration $\tau$ and,~in particular, the logarithmic transformation of schedule duration {$\ln(\tau/\mathrm{dt})$} on~algorithm performance and DD effectiveness. 
Schedule duration reflects the total time required to execute a quantum circuit and depends on the number and execution time of individual gates. Shorter durations potentially improve circuit fidelity by reducing the system's exposure to decoherence errors, but achieving them necessitates faster gates, which can be hardware-limited.

\begin{figure*}[tb]
%
\includegraphics[width=0.48\linewidth]{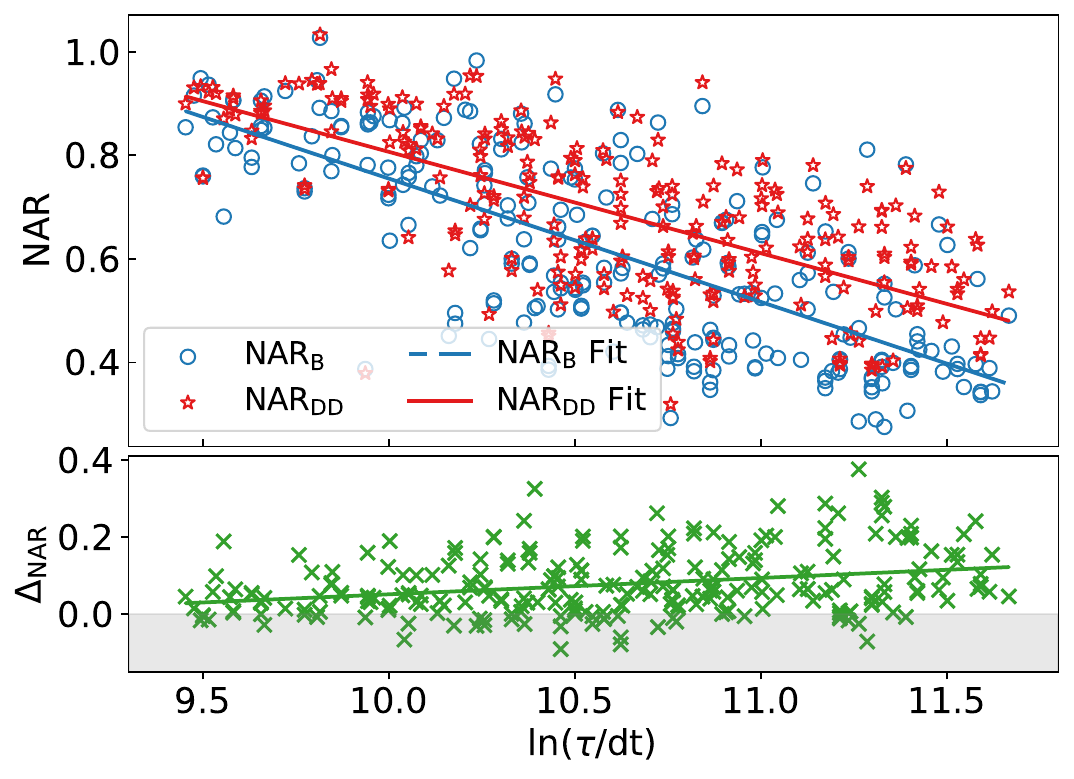}
\put(-0.48\linewidth,0.33\linewidth){(\textbf{a})}
\put(-0.48\linewidth,0.14\linewidth){(\textbf{c})}
\includegraphics[width=0.48\linewidth]{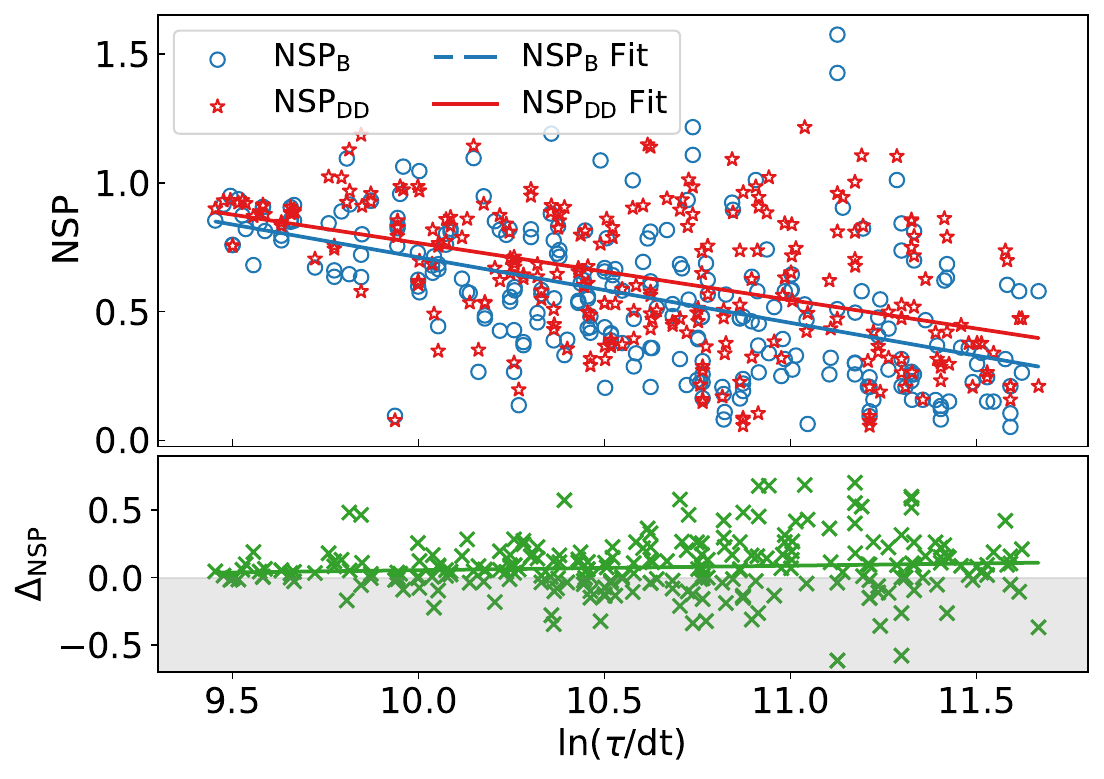}
\put(-0.48\linewidth,0.33\linewidth){(\textbf{b})}
\put(-0.48\linewidth,0.14\linewidth){(\textbf{d})}
\caption{\label{fig:schedule_duration} Influence of logarithmic transformation of circuit schedule duration ({$\ln(\tau/\mathrm{dt})$}) on algorithm performance and DD effectiveness with the same datasets as in Fig.~\ref{fig:circuit_fidelity}: (a) NAR, (b) NSP, (c) $\Delta_\mathrm{\scriptscriptstyle NAR}$, and (d) $\Delta_\mathrm{\scriptscriptstyle NSP}$.}
\end{figure*}

\begin{table*}[tb]
    \caption{\label{tab:sched_dur_para}
    Parameters derived from the analysis of Fig.~\ref{fig:schedule_duration}.}
    \centering
    \begin{ruledtabular}
    
\begin{tabular}{llll}
 \textbf{Metric} & \textbf{Fit Function} & \textbf{Correlation Coefficient} & \textbf{\emph{p}-Value} \\
 \hline
 $\mathrm{NAR}_\mathrm{\scriptscriptstyle B}$ & $y = {-0.238x+3.136}$ &  ${-0.731}$ &  0 \\
 $\mathrm{NAR}_\mathrm{\scriptscriptstyle DD}$ & $y={-0.196x+2.764}$ & ${-0.691}$  & 0 \\
 $\mathrm{NSP}_\mathrm{\scriptscriptstyle B}$ & $y={-0.254x+3.253}$ & ${-0.524}$  &  0 \\
 $\mathrm{NSP}_\mathrm{\scriptscriptstyle DD}$ & $y={-0.221x+2.975}$ & ${-0.464}$  &  0 \\
 \hline
 $\Delta_\mathrm{\scriptscriptstyle NAR}$ & $y=\phantom{-}{0.042x-0.372}$ &  $\phantom{-}{0.295}$ &  0  \\
 $\Delta_\mathrm{\scriptscriptstyle NSP}$ & $y=\phantom{-}{0.033x-0.278}$ &  $\phantom{-}{0.092}$ &   {0.14281}
 \\
\end{tabular}
    \end{ruledtabular}
\end{table*}

Figure~\ref{fig:schedule_duration} depicts the impact of {$\ln(\tau/\mathrm{dt})$} on the defined metrics using the same datasets as in Fig.~\ref{fig:circuit_fidelity}. We observe that the algorithm performance degrades with increasing schedule duration, while DD effectiveness improves. However, $\Delta_\mathrm{\scriptscriptstyle NSP}$ exhibits larger fluctuations for longer durations, suggesting that while DD sequences mitigate decoherence errors, potentially improving performance at longer durations, they could also introduce other error mechanisms, such as operation errors, that counteract this improvement.
Table~\ref{tab:sched_dur_para} summarizes corresponding parameters.
The coefficients of the linear function for $\mathrm{NAR}_\mathrm{\scriptscriptstyle B}$, $\mathrm{NAR}_\mathrm{\scriptscriptstyle DD}$, $\mathrm{NSP}_\mathrm{\scriptscriptstyle B}$, and~$\mathrm{NSP}_\mathrm{\scriptscriptstyle DD}$ indicate a suppressed decay in performance with increasing schedule duration by applying DD sequences.
The correlation coefficients between these metrics and {$\ln(\tau/\mathrm{dt})$} further support the effectiveness of DD sequences in reducing the dependence of performance (both NAR and NSP) on schedule duration.
$\Delta_\mathrm{\scriptscriptstyle NAR}$ exhibits a statistically weak correlation with {$\ln(\tau/\mathrm{dt})$}, while $\Delta_\mathrm{\scriptscriptstyle NSP}$ shows a very weak correlation.
This observation aligns with the findings for circuit fidelity. However, $\Delta_\mathrm{\scriptscriptstyle NSP}$ appears more sensitive to schedule duration compared to circuit fidelity, as~indicated by a larger absolute correlation coefficient ($\abs{C_r}$) and lower $p$-value.

\begin{figure*}[tb]
\includegraphics[width=0.9\linewidth]{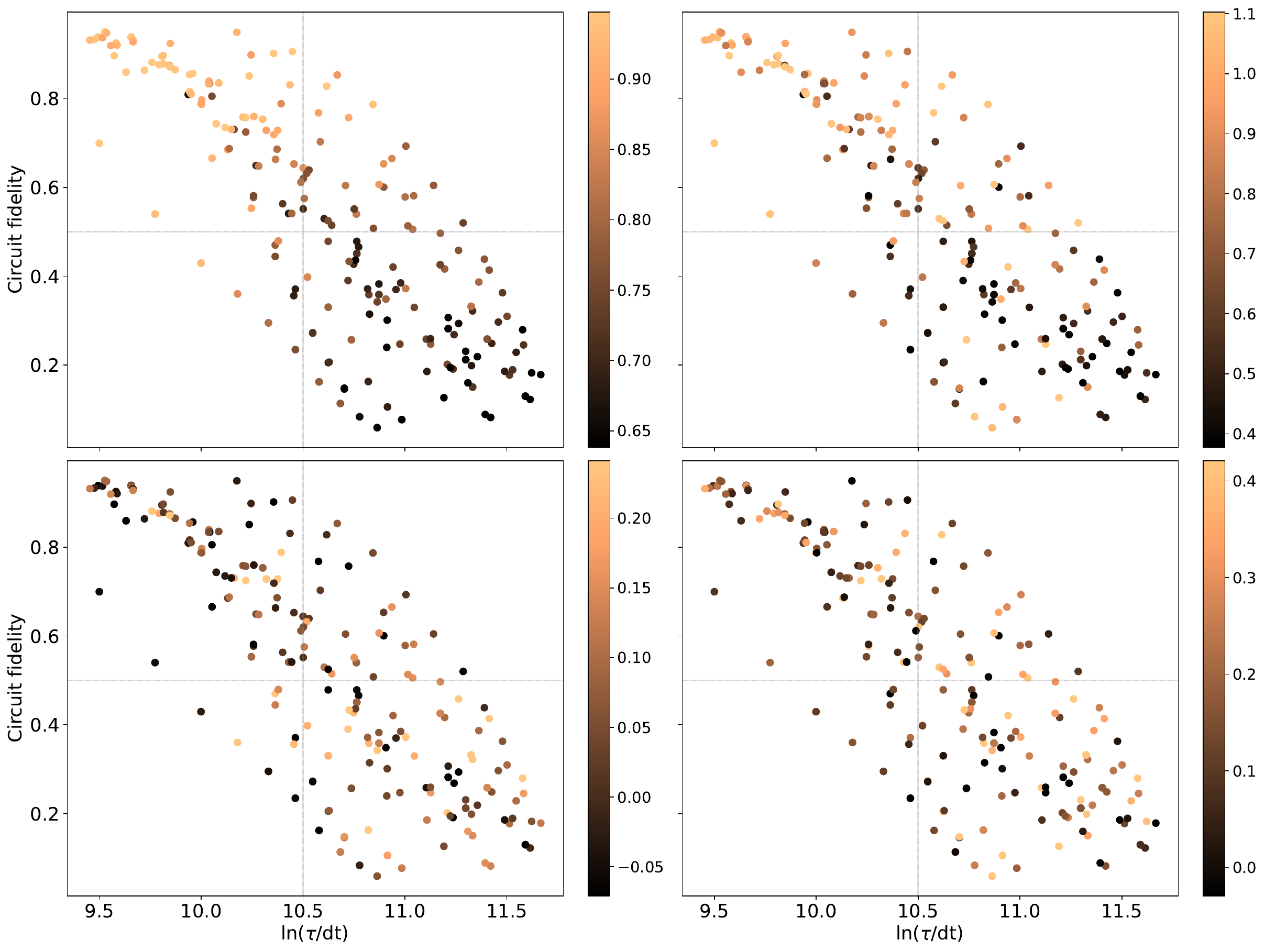}
\put(-0.54\linewidth,0.642\linewidth){{\textbf(a)}}
\put(-0.84\linewidth,0.384\linewidth){\footnotesize $\mathrm{NAR}_\mathrm{DD}$}
\put(-0.108\linewidth,0.642\linewidth){{\textbf(b)}}
\put(-0.396\linewidth,0.384\linewidth){\footnotesize $\mathrm{NSP}_\mathrm{DD}$}
\put(-0.54\linewidth,0.324\linewidth){{\textbf(c)}}
\put(-0.84\linewidth,0.06\linewidth){\footnotesize $\Delta_\mathrm{NAR}$}
\put(-0.108\linewidth,0.324\linewidth){{\textbf(d)}}
\put(-0.396\linewidth,0.06\linewidth){\footnotesize $\Delta_\mathrm{NSP}$}

\caption{\label{fig:schedule_circ_fid_all} 
Influence of circuit fidelity and {$\ln(\tau/\mathrm{dt})$} on algorithm performance and DD effectiveness: (\textbf{a}) $\mathrm{NAR}_\mathrm{\scriptscriptstyle DD}$, (\textbf{b}) $\mathrm{NSP}_\mathrm{\scriptscriptstyle DD}$, (\textbf{c}) $\Delta_\mathrm{\scriptscriptstyle NAR}$, and~(\textbf{d}) $\Delta_\mathrm{\scriptscriptstyle NSP}$.
}
\end{figure*}

Figures~\ref{fig:schedule_circ_fid_all}(a--d) illustrate the impact of circuit fidelity and schedule duration on $\mathrm{NAR}_\mathrm{\scriptscriptstyle DD}$, $\mathrm{NSP}_\mathrm{\scriptscriptstyle DD}$, $\Delta_\mathrm{\scriptscriptstyle NAR}$, and~$\Delta_\mathrm{\scriptscriptstyle NSP}$, respectively.
As observed in Fig.~\ref{fig:schedule_circ_fid_all}(a), $\mathrm{NAR}_\mathrm{\scriptscriptstyle DD}$ exhibits degradation with increasing logarithmic schedule duration ({$\ln(\tau/\mathrm{dt})$}) and decreasing circuit fidelity.
High performance is concentrated in the region where the schedule duration $\tau$ is below {$e^{10.5} \mathrm{dt}$} and circuit fidelity surpasses 0.5. Conversely, low performance is primarily observed for $\tau$ exceeding {$e^{10.5} \mathrm{dt}$} and fidelities below 0.5.
A similar trend is evident for $\mathrm{NSP}_\mathrm{\scriptscriptstyle DD}$ in Fig.~\ref{fig:schedule_circ_fid_all}(b). However, unlike $\mathrm{NAR}_\mathrm{\scriptscriptstyle DD}$, achieving a high $\mathrm{NSP}_\mathrm{\scriptscriptstyle DD}$ value remains feasible even for longer schedule durations and lower fidelities. This suggests that $\mathrm{NSP}_\mathrm{\scriptscriptstyle DD}$ is less sensitive to these factors compared to $\mathrm{NAR}_\mathrm{\scriptscriptstyle DD}$.
Figures~\ref{fig:schedule_circ_fid_all}(c--d) further demonstrate the effectiveness of DD sequences, particularly at longer durations. This is potentially due to the ability of DD sequences to mitigate decoherence errors that become more prominent at these~timescales.

\subsubsection{Native Gate~Sets}

This section explores the performance of quantum algorithms implemented on two distinct sets of IBM QPUs. The~first set comprises four 27-qubit devices utilizing the CX gate as their native two-qubit gate, while the second set consists of four 127-qubit devices employing the ECR gate as their native two-qubit~gate.

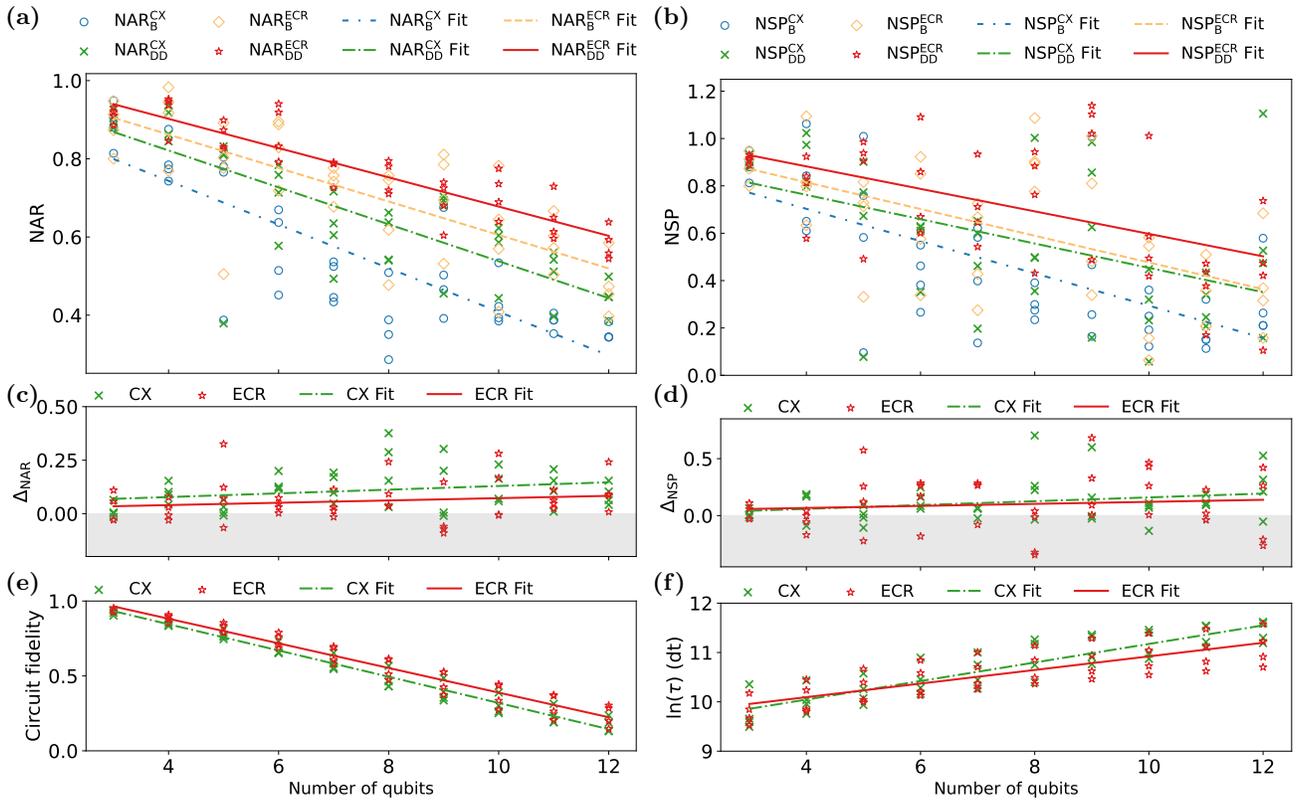
\begin{figure*}[tb]
\includegraphics[width=0.48\linewidth]{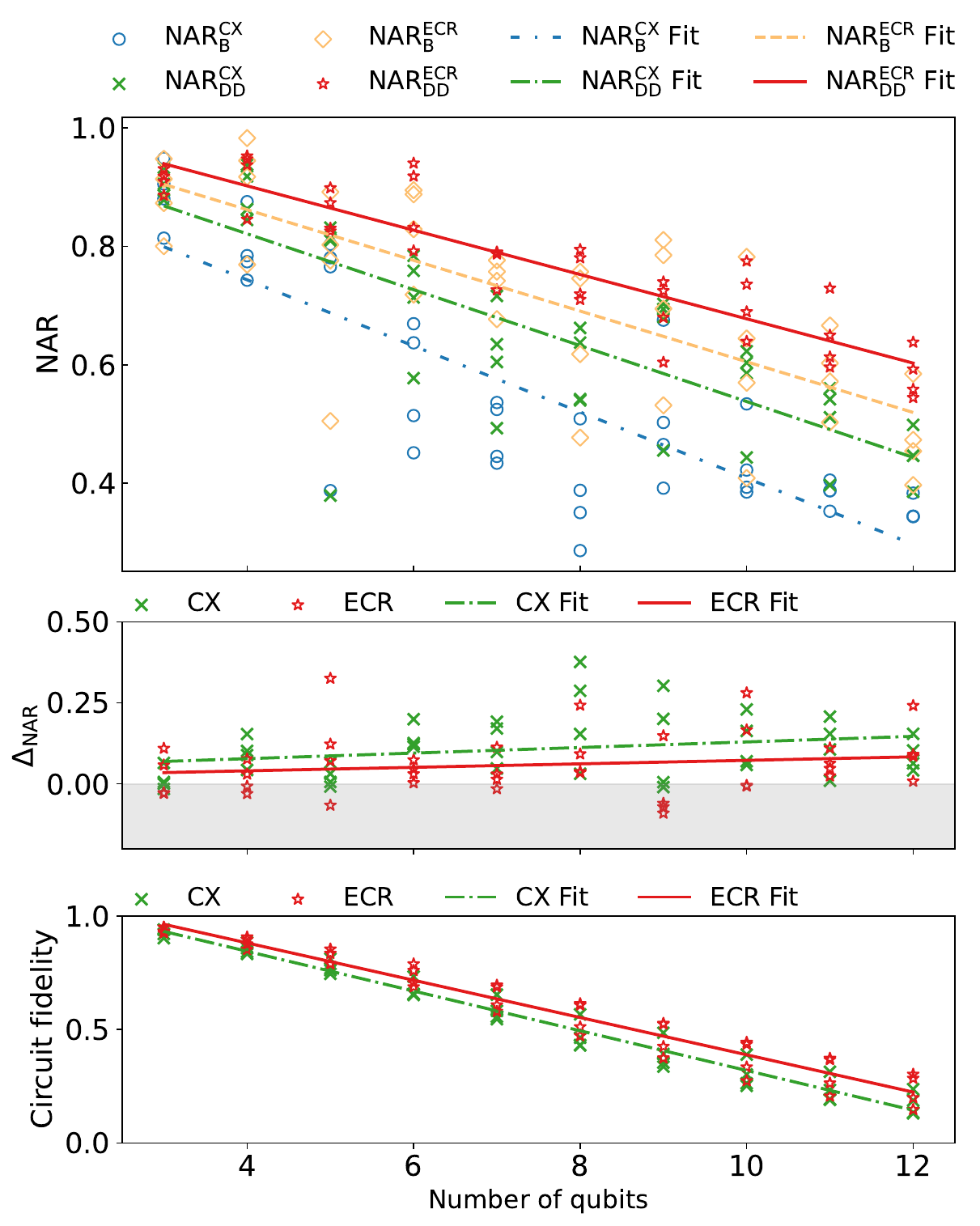}
\put(-0.48\linewidth,0.58\linewidth){(\textbf{a})}
\put(-0.485\linewidth,0.3\linewidth){(\textbf{c})}
\put(-0.48\linewidth,0.16\linewidth){(\textbf{e})}
\includegraphics[width=0.48\linewidth]{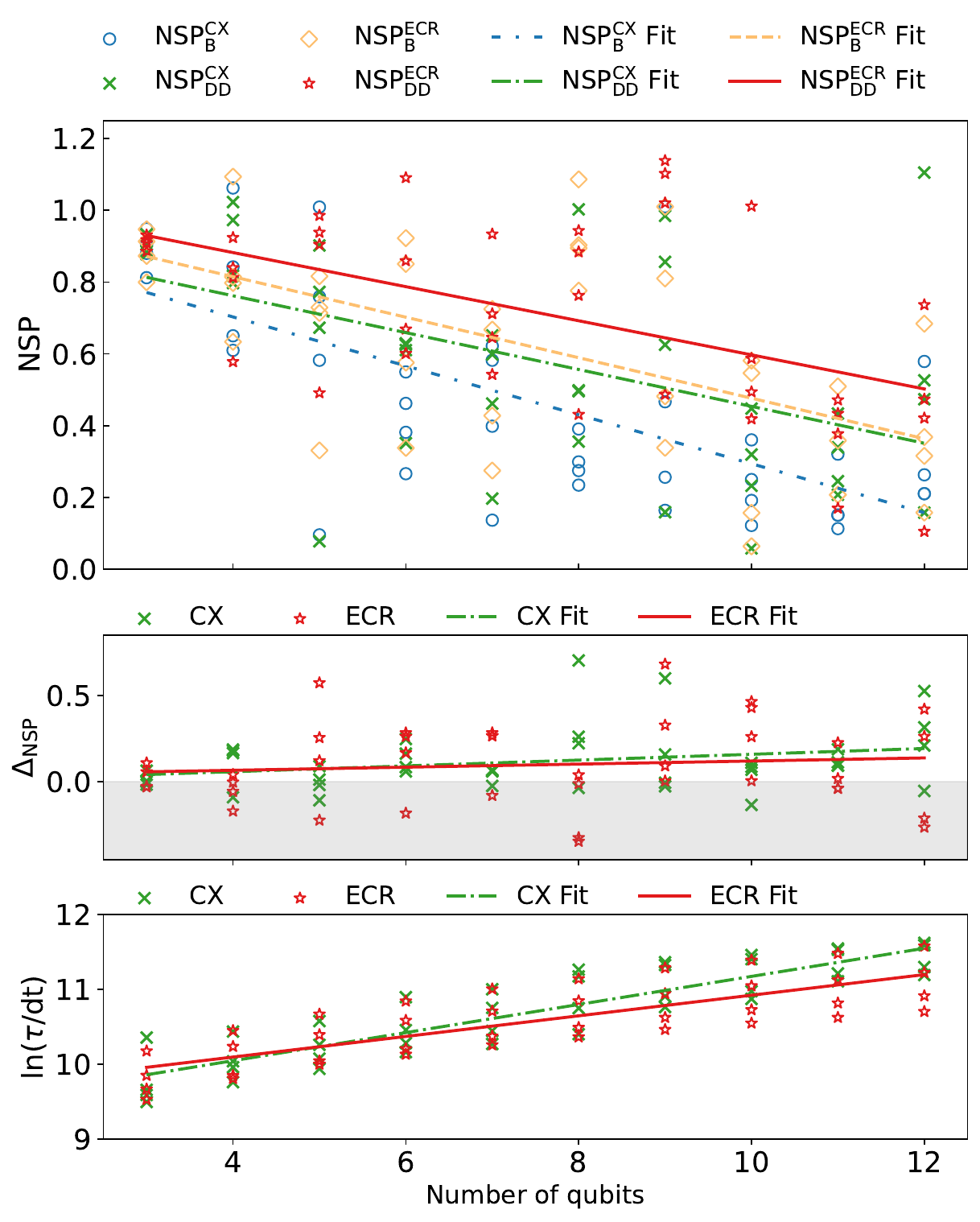}
\put(-0.48\linewidth,0.58\linewidth){(\textbf{b})}
\put(-0.485\linewidth,0.3\linewidth){(\textbf{d})}
\put(-0.48\linewidth,0.16\linewidth){(\textbf{f})}
\caption{\label{fig:native_gate_set}
Comparison of two hardware-native gate sets: \{CX, ID, RZ, SX, X\} and \{ECR, ID, RZ, SX, X\}, denoted as CX and ECR gate sets, respectively. Results obtained using four 27-qubit quantum processing units (QPUs) ibmq\_mumbai, ibmq\_kolkata, ibm\_cairo, and~ibmq\_ehningen for the CX gate set, and~four 127-qubit QPUs ibm\_kyoto, ibm\_cusco, ibm\_brisbane, and~ibm\_sherbrooke for the ECR gate set: (\textbf{a}) NAR, (\textbf{b}) NSP, (\textbf{c}) $\Delta_\mathrm{\scriptscriptstyle NAR}$, (\textbf{d}) $\Delta_\mathrm{\scriptscriptstyle NSP}$, (\textbf{e}) circuit fidelity, and~(\textbf{f}) {$\ln(\tau/\mathrm{dt})$}.
The CPMG sequence is used for all data points. Each line represents a linear fit to the corresponding data.
}
\end{figure*}

Benchmark results obtained from the two QPU sets are presented in Fig.~\ref{fig:native_gate_set}. Table~\ref{tab:native_gate_cx} summarizes average values of metrics and parameters derived from the linear fits of these metrics against the number of qubits. As~shown in Figs.~\ref{fig:native_gate_set}(a) and (b), applying DD sequences generally improves the NAR and NSP, respectively, for~both native gate sets. The~highest performance is achieved with $\mathrm{NAR}_\mathrm{\scriptscriptstyle DD}^\mathrm{\scriptscriptstyle ECR}$ and $\mathrm{NSP}_\mathrm{\scriptscriptstyle DD}^\mathrm{\scriptscriptstyle ECR}$, which leverages ECR-based QPUs and incorporates DD sequences. Moreover, QPUs utilizing the ECR gate exhibit consistently higher baseline performance, $\mathrm{NAR}_\mathrm{\scriptscriptstyle B}^\mathrm{\scriptscriptstyle ECR}$ and $\mathrm{NSP}_\mathrm{\scriptscriptstyle B}^\mathrm{\scriptscriptstyle ECR}$, compared to those with the CX gate, $\mathrm{NAR}_\mathrm{\scriptscriptstyle B}^\mathrm{\scriptscriptstyle CX}$ and $\mathrm{NSP}_\mathrm{\scriptscriptstyle B}^\mathrm{\scriptscriptstyle CX}$, even surpassing those with DD sequences, $\mathrm{NAR}_\mathrm{\scriptscriptstyle DD}^\mathrm{\scriptscriptstyle CX}$, and $\mathrm{NSP}_\mathrm{\scriptscriptstyle DD}^\mathrm{\scriptscriptstyle CX}$. This observation suggests that the selection of QPUs may be more critical for achieving optimal performance than relying solely on DD techniques.
As shown in Figs.~\ref{fig:native_gate_set}(c) and (d), DD effectiveness improves as the qubit count increases for both gate sets. Moreover, the~CX gate set exhibits higher DD effectiveness than the ECR gate set.
{As illustrated in} Figs.~\ref{fig:native_gate_set}(e) and (f), the ECR gate produces an overall higher circuit fidelity and lower schedule duration, potentially contributing to higher performance.
{Analyzing the proportion of positive outcomes in the experimental data presented in Figs.~\ref{fig:native_gate_set}(c) and (d), we observe that the} reported values of $\mathrm{EMSR}_\mathrm{\scriptscriptstyle AR}$ and $\mathrm{EMSR}_\mathrm{\scriptscriptstyle SP}$ are 92.5\% and 72.5\% for the CX gate, respectively, {whereas the} corresponding values are 75\% and 62.5\%, respectively, for~the ECR gate, {suggesting} that DD sequences are more robust in mitigating errors for the CX gate~set.

The correlation coefficient presented in Table~\ref{tab:native_gate_cx} reveals a very strong {negative} correlation between $\mathrm{NAR}_\mathrm{\scriptscriptstyle DD}^\mathrm{\scriptscriptstyle ECR}$ and the number of qubits compared to $\mathrm{NAR}_\mathrm{\scriptscriptstyle B}^\mathrm{\scriptscriptstyle ECR}$. Similarly, a~very strong negative correlation is observed between circuit fidelity and qubit count for both CX and ECR gate sets, indicating a significant decrease in circuit fidelity as the number of qubits increases.
Moreover, we observe a weaker correlation between NSP and qubit number compared to NAR, implying that the impact of qubit count on success probability is less pronounced than its effect on approximation ratio. Additionally, the~high $p$-values for $\Delta_\mathrm{\scriptscriptstyle NAR}^\mathrm{\scriptscriptstyle ECR}$ and $\Delta_\mathrm{\scriptscriptstyle NSP}^\mathrm{\scriptscriptstyle ECR}$ suggest that the effectiveness of DD sequences for ECR gates is more susceptible to random fluctuations. {This can be attributed to the interplay between the intended decoherence suppression capabilities of DD sequences and the additional gate errors that they introduce. As~shown in Figs.~\ref{fig:native_gate_set}(e) and (f), circuits utilizing ECR gates demonstrate higher fidelities while requiring shorter execution times. The~inherent advantage of shorter circuits may limit the potential for further enhancement through the use of DD sequences. In~such cases, incorporating extra DD pulses could even lead to a decrease in overall algorithm performance.}

Our demonstrations on eight IBM QPUs highlight the importance of circuit fidelity, schedule duration, and~DD sequences in optimizing algorithm performance. As~circuit fidelity decreases and schedule duration increases, DD sequences become increasingly important for mitigating errors and identifying optimal solutions.
Furthermore, the~results suggest that ECR-based QPUs offer advantages over CX-based QPUs. This is primarily due to the inherently shorter schedule durations and higher circuit fidelities associated with ECR gates. However, CX-based QPUs benefit more significantly from DD sequences in terms of error~mitigation.

\begin{table*}[tb]
    \caption{\label{tab:native_gate_cx}
    Parameters derived from the analysis of Fig.~\ref{fig:native_gate_set}.
    }
    \centering
    \begin{ruledtabular}
\begin{tabular}{lllll}
         \textbf{Metric} & \textbf{Mean} & \textbf{Fit Function} & \textbf{Correlation Coefficient} & \textbf{\emph{p}-Value} \\
         \hline
         $\mathrm{NAR}_\mathrm{\scriptscriptstyle B}^\mathrm{\scriptscriptstyle CX}$ & 0.548 & $y=-0.056x+0.967$ &  $-0.832$ &  0 \\
         $\mathrm{NAR}_\mathrm{\scriptscriptstyle DD}^\mathrm{\scriptscriptstyle CX}$ & 0.656 & $y=-0.047x+1.010$ & $-0.812$  & 0 \\
         $\mathrm{NAR}_\mathrm{\scriptscriptstyle B}^\mathrm{\scriptscriptstyle ECR}$ & 0.712 & $y=-0.043x+1.033$ &  $-0.766$ &  0 \\
         $\mathrm{NAR}_\mathrm{\scriptscriptstyle DD}^\mathrm{\scriptscriptstyle ECR}$ & 0.771 & $y=-0.037x+1.052$ & $-0.911$  & 0 \\
         \hline
         $\mathrm{NSP}_\mathrm{\scriptscriptstyle B}^\mathrm{\scriptscriptstyle CX}$ & 0.464 & $y=-0.068x+0.976$ &  $-0.678$ &  0 \\
         $\mathrm{NSP}_\mathrm{\scriptscriptstyle DD}^\mathrm{\scriptscriptstyle CX}$ & 0.582 & $y=-0.051x+0.967$ & $-0.504$  & 0.00092 \\
         $\mathrm{NSP}_\mathrm{\scriptscriptstyle B}^\mathrm{\scriptscriptstyle ECR}$ & 0.618 & $y=-0.056x+1.041$ &  $-0.579$ &  0.00009\\
         $\mathrm{NSP}_\mathrm{\scriptscriptstyle DD}^\mathrm{\scriptscriptstyle ECR}$ & 0.716 & $y=-0.047x+1.072$ & $-0.529$  & 0.00044 \\
         
         \hline
         $\Delta_\mathrm{\scriptscriptstyle NAR}^\mathrm{\scriptscriptstyle CX}$ & 0.108 & $y=\phantom{-}0.009x+0.043$ &  $\phantom{-}0.27$ &  0.09252  \\
         $\Delta_\mathrm{\scriptscriptstyle NAR}^\mathrm{\scriptscriptstyle ECR}$ & 0.059 & $y=\phantom{-}0.005x+0.019$ &  $\phantom{-}0.167$ & 0.30429  \\
         $\Delta_\mathrm{\scriptscriptstyle NSP}^\mathrm{\scriptscriptstyle CX}$ & 0.118 & $y=\phantom{-}0.017x-0.008$ &  $\phantom{-}0.277$ &  0.08407  \\
         $\Delta_\mathrm{\scriptscriptstyle NSP}^\mathrm{\scriptscriptstyle ECR}$ & 0.098 & $y=\phantom{-}0.009x+0.031$ &  $\phantom{-}0.108$ & 0.50688  \\
         \hline
         Circuit fidelity (CX) & 0.539 & $y=-0.088x+1.195$ &  $-0.984$ & 0  \\
         Circuit fidelity (ECR) & 0.594 & $y=-0.082x+1.211$ &  $-0.975$ & 0  \\
         {$\ln(\tau/\mathrm{dt})$} (CX) & 10.703 & $y=\phantom{-}0.188x+9.294$ &  $\phantom{-}0.891$ & 0  \\
         {$\ln(\tau/\mathrm{dt})$} (ECR) & 10.577 & $y=\phantom{-}0.138x+9.543$ &  $\phantom{-}0.791$ & 0  \\
\end{tabular}
    \end{ruledtabular}

\end{table*}

\subsection{Impact of Algorithmic~Factors}

This section investigates the influence of algorithmic factors on the algorithm performance and DD effectiveness. We focus on three key aspects: algorithm implementations, DD sequence types, and~circuit optimization~levels.

\subsubsection{Algorithm~Implementations \label{subsubsec:algor_imple}}

\begin{figure*}[tb]
\includegraphics[width=0.48\linewidth]{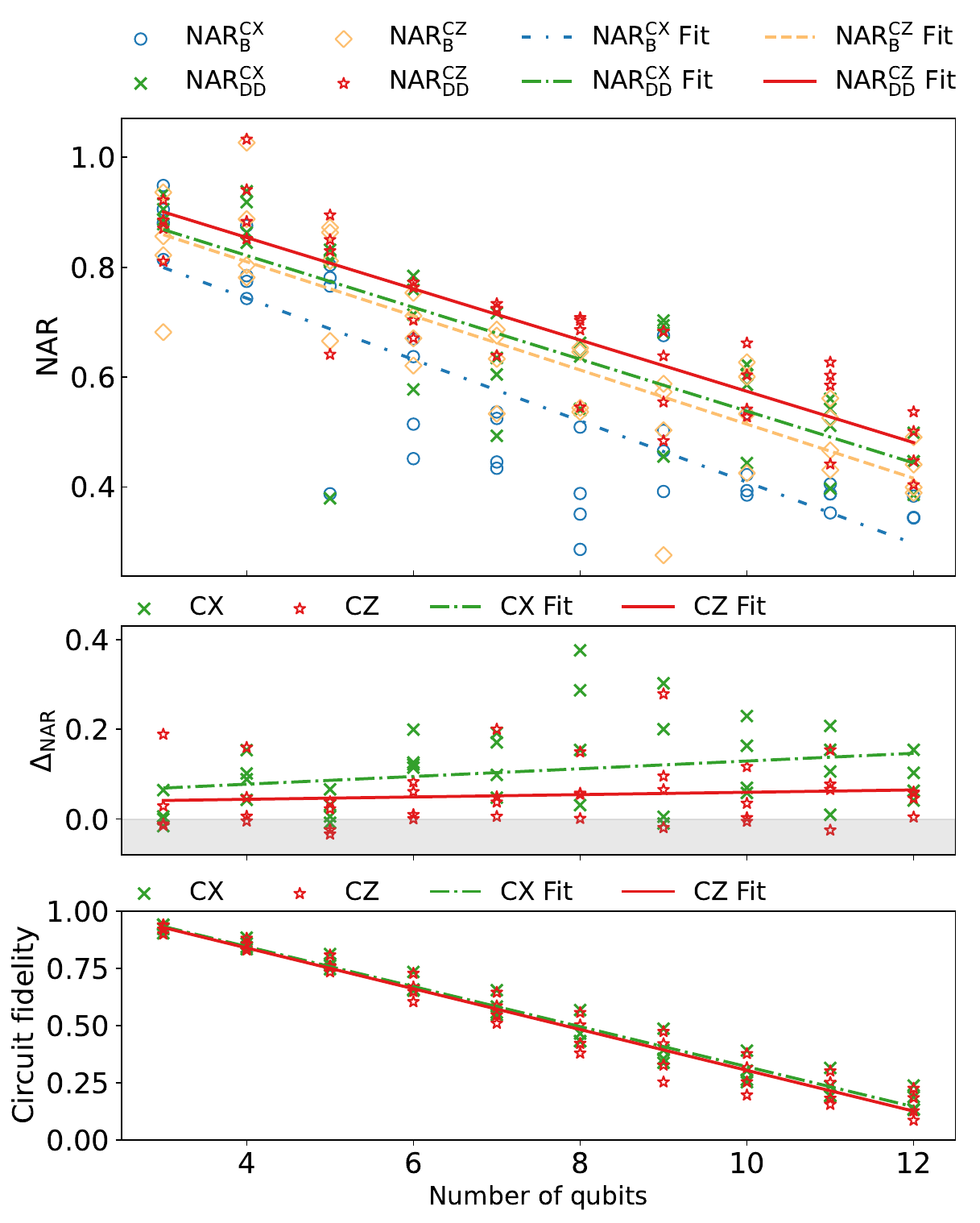}
\put(-0.48\linewidth,0.58\linewidth){(\textbf{a})}
\put(-0.48\linewidth,0.3\linewidth){(\textbf{c})}
\put(-0.48\linewidth,0.16\linewidth){(\textbf{e})}
\includegraphics[width=0.48\linewidth]{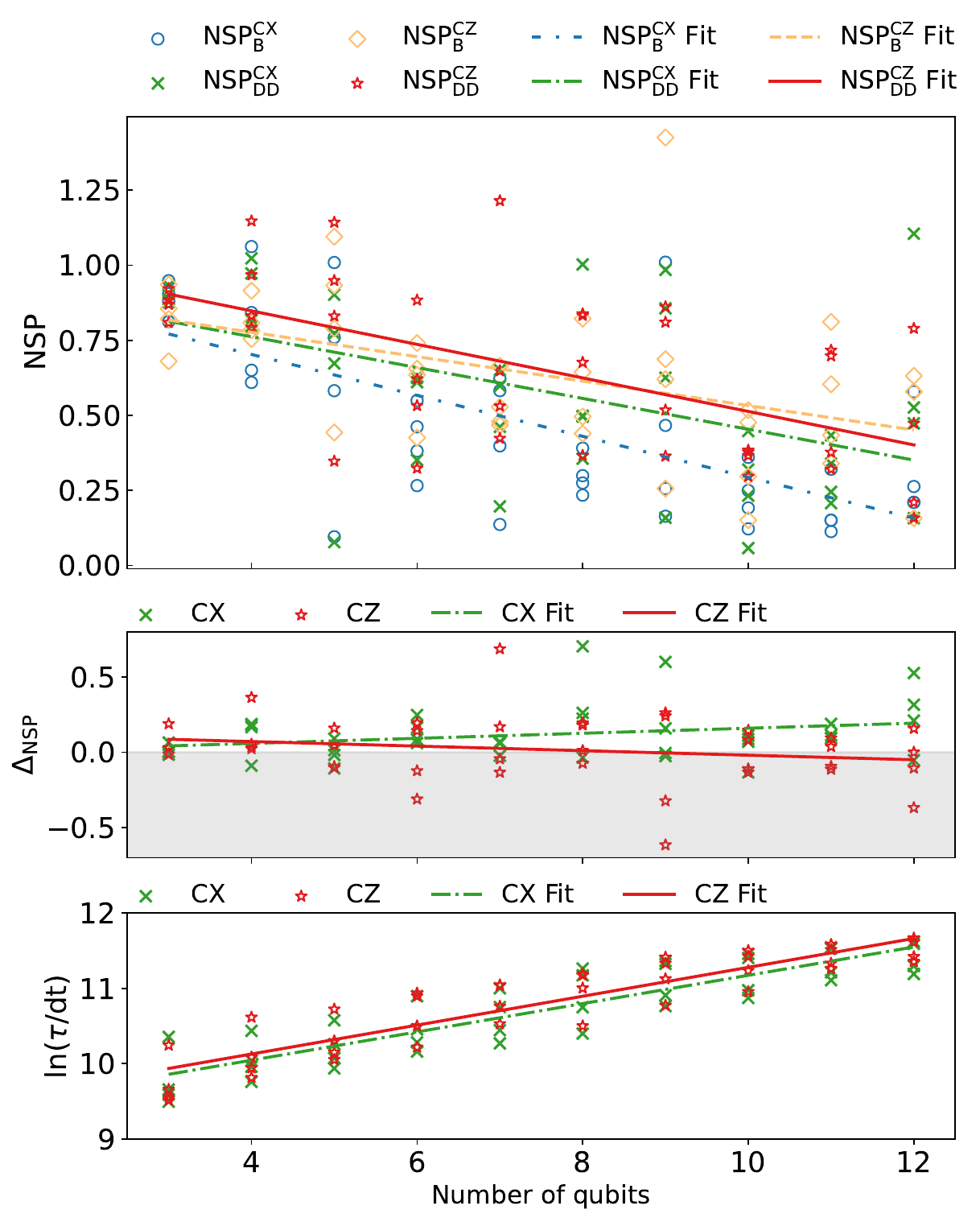}
\put(-0.48\linewidth,0.58\linewidth){(\textbf{b})}
\put(-0.48\linewidth,0.3\linewidth){(\textbf{d})}
\put(-0.48\linewidth,0.16\linewidth){(\textbf{f})}
\caption{\label{fig:algorithm_implementation}
Comparison of CX and CZ implementations of QAOA across four 27-qubit IBM QPUs ibmq\_mumbai, ibmq\_kolkata, ibm\_cairo, and~ibmq\_ehningen: (\textbf{a}) NAR, (\textbf{b}) NSP, (\textbf{c}) $\Delta_\mathrm{\scriptscriptstyle NAR}$, (\textbf{d}) $\Delta_\mathrm{\scriptscriptstyle NSP}$, (\textbf{e}) circuit fidelity, and~(\textbf{f}) {$\ln(\tau/\mathrm{dt})$}.
The CPMG sequence is used for all data points.
Each line represents a linear fit of the data.}
\end{figure*}

We compare the performance of CX and CZ implementations of the QAOA on four 27-qubit IBM QPUs, as~detailed in Section~\ref{subsec:syner_design_appro}. The~CPMG sequence is consistently utilized throughout this analysis.
As shown in Figs.~\ref{fig:algorithm_implementation}(a) and (b), CZ implementation with DD sequences achieves the highest average values for both the NAR ($\mathrm{NAR}_\mathrm{\scriptscriptstyle DD}^\mathrm{\scriptscriptstyle CZ}$) and NSP ($\mathrm{NSP}_\mathrm{\scriptscriptstyle DD}^\mathrm{\scriptscriptstyle CZ}$). However, for~the NSP at qubit counts exceeding 9, CZ implementation without DD sequences, $\mathrm{NSP}_\mathrm{\scriptscriptstyle B}^\mathrm{\scriptscriptstyle CZ}$, outperforms that with DD sequences, $\mathrm{NSP}_\mathrm{\scriptscriptstyle DD}^\mathrm{\scriptscriptstyle CZ}$, potentially due to the introduction of significant errors by DD sequences themselves.
Furthermore, CX implementation exhibits increasing DD effectiveness as~the qubit number grows (Figs.~\ref{fig:algorithm_implementation}(c) and (d)){, whereas} $\Delta_\mathrm{\scriptscriptstyle NSP}$ for CZ implementation {exhibits a slight decrease}. Additionally, $\mathrm{EMSR}_\mathrm{\scriptscriptstyle AR}$ and $\mathrm{EMSR}_\mathrm{\scriptscriptstyle SP}$ are consistently higher for CX implementation (92.5\% and 80\%, respectively) compared to CZ implementation (75\% and 57.5\%, respectively), suggesting a higher robustness of DD sequences for CX implementation.
{As depicted in Figs.~\ref{fig:algorithm_implementation}(e) and (b),} CX and CZ implementations exhibit comparable circuit fidelities. For~a larger number of qubits, CX implementation even achieves slightly higher fidelities. Moreover, CX implementation demonstrates a consistently shorter schedule duration compared to CZ implementation {(Fig.~\ref{fig:algorithm_implementation}(f))}. This difference in schedule duration is attributed to the increased number of single-qubit gates required by CZ implementation (details in Fig.~\ref{fig:circuit_algorithm_implementations}).

Table~\ref{tab:algor_imp_cx_cz} summarizes the average value of each metric along with the linear fit parameters extracted from Fig.~\ref{fig:algorithm_implementation}.
The negative coefficient associated with $\Delta_\mathrm{\scriptscriptstyle NSP}^\mathrm{\scriptscriptstyle CZ}$ suggests a decrease in DD effectiveness, {as measured by success probability,} with an increasing qubit count for CZ implementation.
{This behavior can be attributed to the intricate interplay between the optimization landscape and gate errors. As~the number of qubits involved increases, the~cumulative error introduced by DD sequences becomes more pronounced. This significantly alters the optimization landscape, rendering the pre-optimized parameters of the QAOA in the noiseless case no longer suitable.}
Furthermore, correlation coefficients between the NAR ($\mathrm{NAR}_\mathrm{\scriptscriptstyle B}$ and $\mathrm{NAR}_\mathrm{\scriptscriptstyle DD}$) and qubit count reveal that applying DD sequences weakens the correlation for CX implementation while strengthening it for CZ implementation.
This trend is also observed for $\mathrm{NSP}_\mathrm{\scriptscriptstyle B}$ and $\mathrm{NSP}_\mathrm{\scriptscriptstyle DD}$.
An additional observation is the high $p$-values associated with $\Delta_\mathrm{\scriptscriptstyle NAR}^\mathrm{\scriptscriptstyle CZ}$ and $\Delta_\mathrm{\scriptscriptstyle NSP}^\mathrm{\scriptscriptstyle CZ}$, indicating the potential dominance of random fluctuations in these metrics for CZ~implementation.

The results suggest that although CX implementation offers more advantages in terms of DD effectiveness, the~higher performance of CZ implementation highlights the significance of circuit structure in executing the QAOA. In~certain instances, the~inherent advantage of a more symmetrical circuit structure, as~exhibited by CZ implementation, can outweigh the benefits of strong DD mitigation achieved with CX implementation. However, the~optimal selection of gate decomposition ultimately relies on the specific algorithm being implemented, the~hardware capabilities available, and~the desired balance between overhead caused by DD and potential performance~gains.

\begin{table*}[tb]
    \caption{\label{tab:algor_imp_cx_cz}
    Parameters derived from the analysis of Fig.~\ref{fig:algorithm_implementation}.
    }
    \centering
    \begin{ruledtabular}
    
\begin{tabular}{lllll}
 \textbf{Metric} & \textbf{Mean} & \textbf{Fit Function} & \textbf{Correlation Coefficient} & \textbf{\emph{p}-Value} \\
 \hline
 $\mathrm{NAR}_\mathrm{\scriptscriptstyle B}^\mathrm{\scriptscriptstyle CX}$ & 0.548 & $y=-0.056x+0.967$ &  $-0.832$ &  0 \\
 $\mathrm{NAR}_\mathrm{\scriptscriptstyle DD}^\mathrm{\scriptscriptstyle CX}$ & 0.656 & $y=-0.047x+1.010$ & $-0.812$  & 0 \\
 $\mathrm{NAR}_\mathrm{\scriptscriptstyle B}^\mathrm{\scriptscriptstyle CZ}$ & 0.638 & $y=-0.049x+1.007$ &  $-0.850$ &  0 \\
 $\mathrm{NAR}_\mathrm{\scriptscriptstyle DD}^\mathrm{\scriptscriptstyle CZ}$ & 0.691 & $y=-0.047x+1.041$ & $-0.885$  & 0 \\
 \hline
 $\mathrm{NSP}_\mathrm{\scriptscriptstyle B}^\mathrm{\scriptscriptstyle CX}$ & 0.464 & $y=-0.068x+0.976$ &  $-0.678$ &  0 \\
 $\mathrm{NSP}_\mathrm{\scriptscriptstyle DD}^\mathrm{\scriptscriptstyle CX}$ & 0.582 & $y=-0.051x+0.967$ & $-0.504$  & 0.00092 \\
 $\mathrm{NSP}_\mathrm{\scriptscriptstyle B}^\mathrm{\scriptscriptstyle CZ}$ & 0.635 & $y=-0.041x+0.940$ &  $-0.473$ &  0.00208\\
 $\mathrm{NSP}_\mathrm{\scriptscriptstyle DD}^\mathrm{\scriptscriptstyle CZ}$ & 0.653 & $y=-0.056x+1.071$ & $-0.589$  & 0.00006 \\
 
 \hline
 $\Delta_\mathrm{\scriptscriptstyle NAR}^\mathrm{\scriptscriptstyle CX}$ & 0.108 & $y=\phantom{-}0.009x+0.043$ &  $\phantom{-}0.27$ &  0.09252  \\
 $\Delta_\mathrm{\scriptscriptstyle NAR}^\mathrm{\scriptscriptstyle CZ}$ & 0.053 & $y=\phantom{-}0.003x+0.034$ &  $\phantom{-}0.109$ & 0.50206  \\
 $\Delta_\mathrm{\scriptscriptstyle NSP}^\mathrm{\scriptscriptstyle CX}$ & 0.118 & $y=\phantom{-}0.017x-0.008$ &  $\phantom{-}0.277$ &  0.08407  \\
 $\Delta_\mathrm{\scriptscriptstyle NSP}^\mathrm{\scriptscriptstyle CZ}$ & 0.018 & $y=-0.015x+0.132$ &  $-0.202$ & 0.21130  \\
 \hline
 Circuit fidelity (CX) & 0.539 & $y=-0.088x+1.195$ &  $-0.984$ & 0  \\
 Circuit fidelity (CZ) & 0.527 & $y=-0.089x+1.195$ &  $-0.977$ & 0  \\
 {$\ln(\tau/\mathrm{dt})$} (CX) & 10.703 & $y=\phantom{-}0.188x+9.294$ &  $\phantom{-}0.891$ & 0  \\
 {$\ln(\tau/\mathrm{dt})$} (CZ) & 10.799 & $y=\phantom{-}0.192x+9.360$ &  $\phantom{-}0.901$ & 0  \\
\end{tabular}
    \end{ruledtabular}

\end{table*}

\subsubsection{DD~Sequences}

This section evaluates the performance of two DD sequences, CPMG and XY4, for~mitigating decoherence errors during QAOA execution with CX implementation.
The evaluation leverages data from seven IBM QPUs. XY4, which employs four single-qubit pulses, might be more effective for qubits with extended idle times than CPMG, which utilizes two single-qubit pulses. {While DD sequences may introduce crosstalk errors, the~use of the same transpiled circuits minimizes the impact of this potential crosstalk on our evaluation.}

\begin{figure*}[tb]
\includegraphics[width=0.48\linewidth]{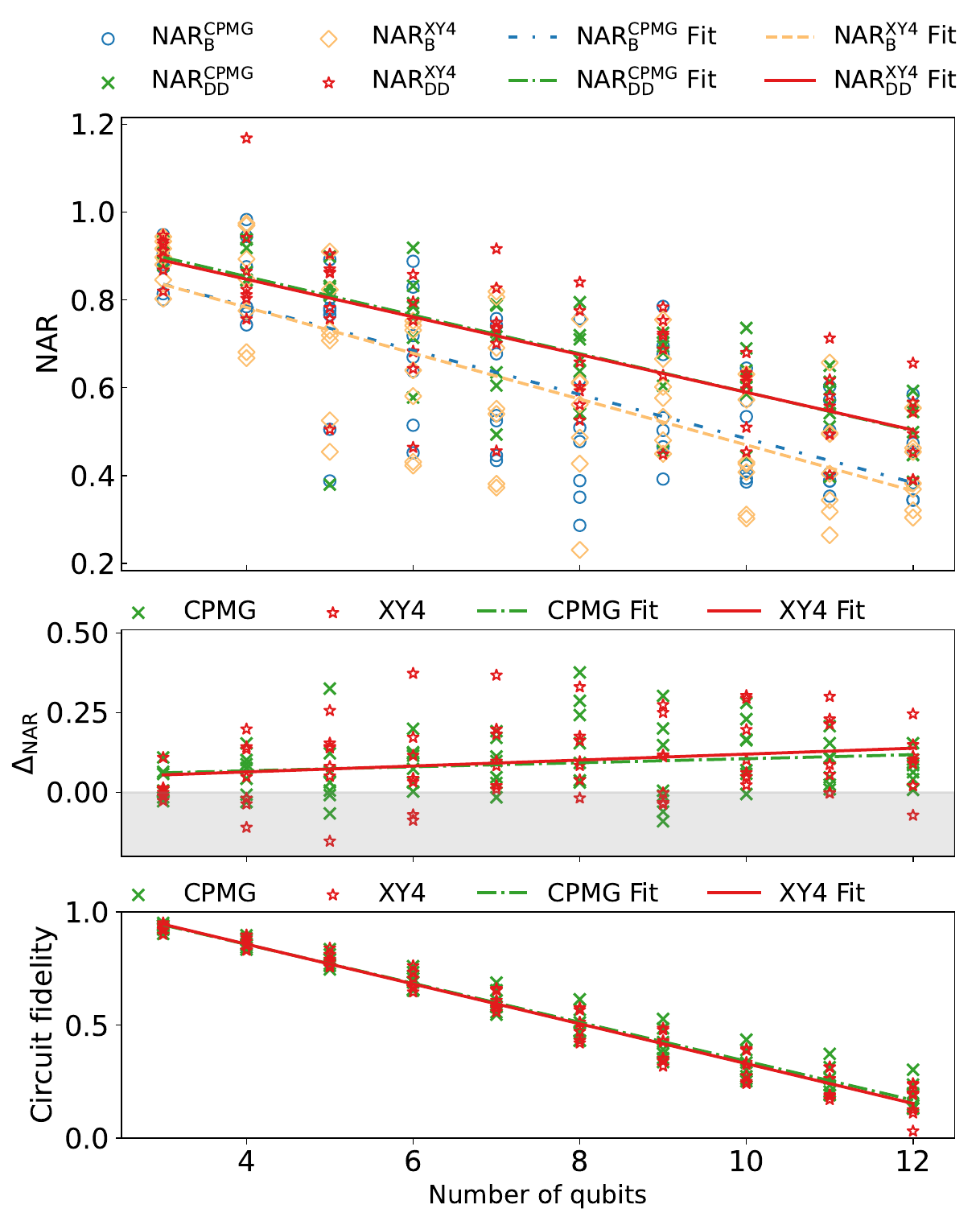}
\put(-0.48\linewidth,0.58\linewidth){(\textbf{a})}
\put(-0.48\linewidth,0.3\linewidth){(\textbf{c})}
\put(-0.48\linewidth,0.16\linewidth){(\textbf{e})}
\includegraphics[width=0.48\linewidth]{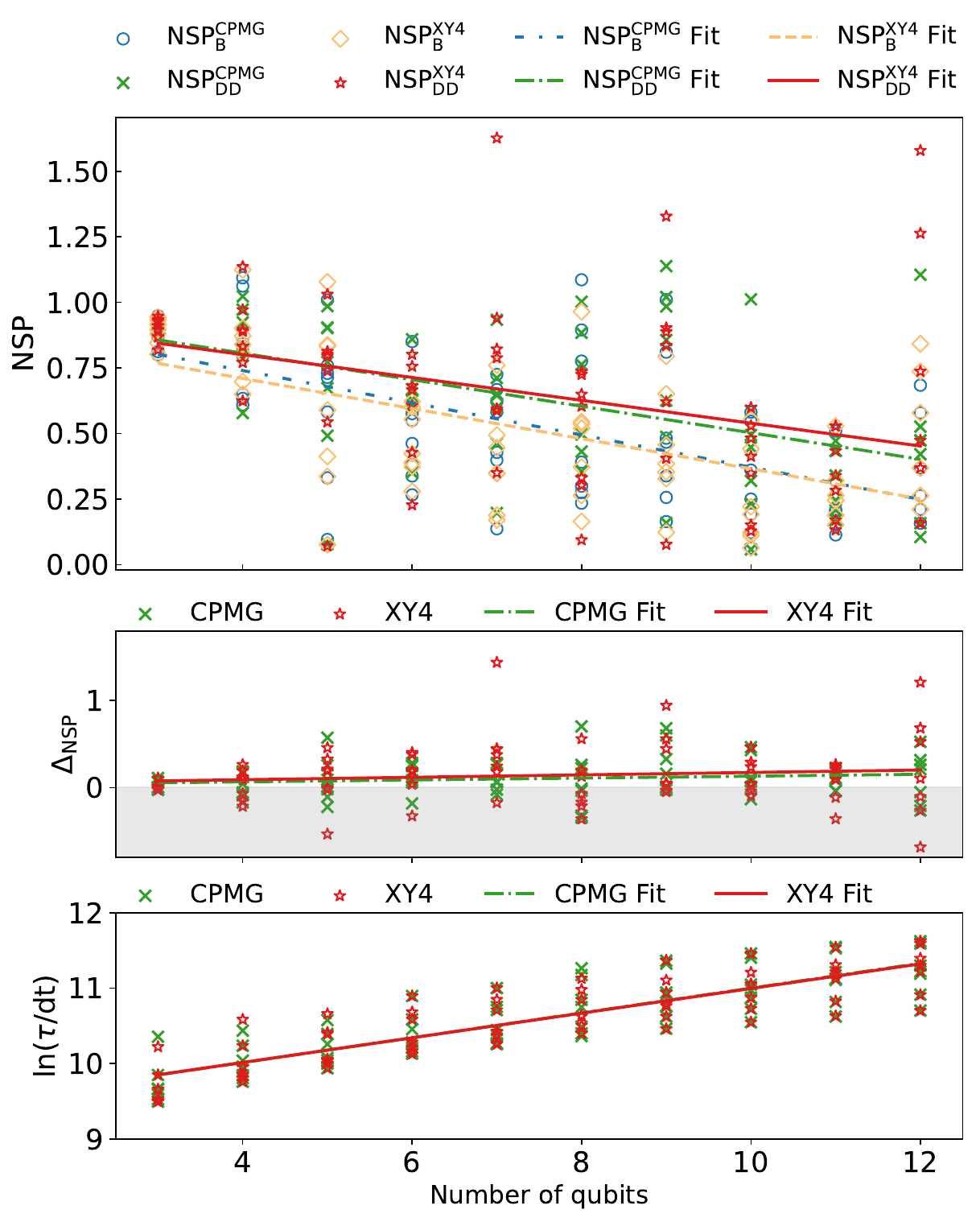}
\put(-0.48\linewidth,0.58\linewidth){(\textbf{b})}
\put(-0.48\linewidth,0.3\linewidth){(\textbf{d})}
\put(-0.48\linewidth,0.16\linewidth){(\textbf{f})}
\caption{\label{fig:dd_pulse_sequence}
Comparison of CPMG and XY4 sequences across seven IBM QPUs ibmq\_mumbai, ibmq\_kolkata, ibm\_cairo, ibmq\_ehningen, ibm\_kyoto, ibm\_cusco, and~ibm\_brisbane: (\textbf{a}) NAR, (\textbf{b}) NSP, (\textbf{c}) $\Delta_\mathrm{\scriptscriptstyle NAR}$, (\textbf{d}) $\Delta_\mathrm{\scriptscriptstyle NSP}$, (\textbf{e}) circuit fidelity, and~(\textbf{f}) {$\ln(\tau/\mathrm{dt})$}. Each line represents a linear fit of the data.
}
\end{figure*}
\begin{table*}[tb]
    \caption{\label{tab:dd_pulse_sequence} Parameters derived from the analysis of Fig.~\ref{fig:dd_pulse_sequence}.}
    \centering
    \begin{ruledtabular}
    
\begin{tabular}{lllll}
     \textbf{Metric} & \textbf{Mean} & \textbf{Fit Function} & \textbf{Correlation Coefficient} & \textbf{\emph{p}-Value} \\
     \hline
     $\mathrm{NAR}_\mathrm{\scriptscriptstyle B}^\mathrm{\scriptscriptstyle CPMG}$ & 0.610 & $y=-0.050x+0.986$ &  $-0.749$ &  0 \\
     $\mathrm{NAR}_\mathrm{\scriptscriptstyle DD}^\mathrm{\scriptscriptstyle CPMG}$ & 0.699 & $y=-0.044x+1.029$ & $-0.802$  & 0 \\
     $\mathrm{NAR}_\mathrm{\scriptscriptstyle B}^\mathrm{\scriptscriptstyle XY4}$ & 0.6 & $y=-0.052x+0.992$ &  $-0.752$ &  0 \\
     $\mathrm{NAR}_\mathrm{\scriptscriptstyle DD}^\mathrm{\scriptscriptstyle XY4}$ & 0.697 & $y=-0.043x+1.019$ & $-0.750$  & 0 \\
     \hline
     $\mathrm{NSP}_\mathrm{\scriptscriptstyle B}^\mathrm{\scriptscriptstyle CPMG}$ & 0.525 & $y=-0.061x+0.986$ &  $-0.613$ &  0 \\
     $\mathrm{NSP}_\mathrm{\scriptscriptstyle DD}^\mathrm{\scriptscriptstyle CPMG}$ & 0.629 & $y=-0.051x+1.009$ & $-0.513$  & 0.00001 \\
     $\mathrm{NSP}_\mathrm{\scriptscriptstyle B}^\mathrm{\scriptscriptstyle XY4}$ & 0.509 & $y=-0.057x+0.940$ &  $-0.591$ &  0\\
     $\mathrm{NSP}_\mathrm{\scriptscriptstyle DD}^\mathrm{\scriptscriptstyle XY4}$ & 0.648 & $y=-0.044x+0.975$ & $-0.373$  & 0.00148 \\
     
     \hline
     $\Delta_\mathrm{\scriptscriptstyle NAR}^\mathrm{\scriptscriptstyle CPMG}$ & 0.09 & $y=\phantom{-}0.006x+0.042$ &  $\phantom{-}0.189$ &  0.11748  \\
     $\Delta_\mathrm{\scriptscriptstyle NAR}^\mathrm{\scriptscriptstyle XY4}$ & 0.097 & $y=\phantom{-}0.009x+0.027$ &  $\phantom{-}0.23$ & 0.05502 \\
     $\Delta_\mathrm{\scriptscriptstyle NSP}^\mathrm{\scriptscriptstyle CPMG}$ & 0.104 & $y=\phantom{-}0.011x+0.023$ &  $\phantom{-}0.144$ &  0.23425  \\
     $\Delta_\mathrm{\scriptscriptstyle NSP}^\mathrm{\scriptscriptstyle XY4}$ & 0.139 & $y=\phantom{-}0.014x+0.035$ &  $\phantom{-}0.116$ & 0.33818  \\
     \hline
     Circuit fidelity (CPMG) & 0.555 & $y=-0.086x+1.201$ &  $-0.979$ & 0  \\
     Circuit fidelity (XY4) & 0.55 & $y=-0.088x+1.209$ &  $-0.981$ & 0  \\
     {$\ln(\tau/\mathrm{dt})$} (CPMG) & 10.589 & $y=\phantom{-}0.165x+9.354$ &  $\phantom{-}0.851$ & 0  \\
     {$\ln(\tau/\mathrm{dt})$} (XY4) & 10.586 & $y=\phantom{-}0.163x+9.360$ &  $\phantom{-}0.858$ & 0  \\
\end{tabular}
    \end{ruledtabular}

\end{table*}

Figures~\ref{fig:dd_pulse_sequence}(a) and (b) {show} that both CPMG and XY4 contribute to improved algorithm performance. While CPMG and XY4 achieve comparable NAR values, XY4 exhibits a better NSP for a larger number of qubits{, but~with noticeable fluctuations}.
{CPMG and XY4 demonstrate comparable DD effectiveness for a small number of qubits (Figs.~\ref{fig:dd_pulse_sequence}(c) and (d)), with~XY4 showing a slight advantage for larger qubit counts.} {However, CPMG exhibits greater robustness, as~evidenced by its higher $\mathrm{EMSR}_\mathrm{\scriptscriptstyle AR}$ (84.29\%) and $\mathrm{EMSR}_\mathrm{\scriptscriptstyle SP}$ (75.71\%) compared to XY4's values (67.14\% and 64.29\%, respectively).}
Figures~\ref{fig:dd_pulse_sequence}(e) and (f) illustrate {comparable} circuit fidelity and schedule duration for both CPMG and XY4.
As before, the~data are fitted with a linear function. The~resulting parameters are summarized in Table~\ref{tab:dd_pulse_sequence}. The~coefficients of the linear function suggest that both CPMG and XY4 effectively suppress the decrease in NAR and NSP as the qubit count increases. The~correlation coefficients and $p$-values suggest a stronger correlation between $\Delta_\mathrm{\scriptscriptstyle NAR}^\mathrm{\scriptscriptstyle XY4}$ and qubit count compared to $\Delta_\mathrm{\scriptscriptstyle NAR}^\mathrm{\scriptscriptstyle CPMG}$, $\Delta_\mathrm{\scriptscriptstyle NSP}^\mathrm{\scriptscriptstyle XY4}$, and~$\Delta_\mathrm{\scriptscriptstyle NSP}^\mathrm{\scriptscriptstyle CPMG}$.

The results indicate that DD {sequences are} generally recommended {for improving circuit performance. They allow for achieving acceptable results in a wider range of circuits. For~instance, the QAOA with DD can reach higher NAR values for more qubits.} However, the~effectiveness and robustness of DD sequences can vary. While XY4 {offers slightly better performance improvements in terms of NAR and NSP}, CPMG demonstrates higher robustness as {measured by EMSR}. This highlights the importance of considering {both performance gains and mitigation robustness when choosing} a DD~sequence.

\subsubsection{Optimization~Levels}

We investigate the influence of optimization levels on the performance and DD effectiveness using five IBM QPUs. Two optimization levels {within Qiskit's transpiler} are considered: level 1 (Opt1) and level 3 (Opt3), representing the default and highest settings, respectively. It is worth noting that different optimization levels do not affect the number of two-qubit gates in our demonstration. This is because we are considering benchmark circuits that have already undergone the AOQMAP approach~\cite{ji2023algorithm}, which effectively ensures adherence to connectivity constraints and eliminates the need for additional SWAP gates. In~our case, different optimization levels influence the selected qubits for circuit execution and the count of single-qubit gates, which affects the circuit fidelity and schedule duration. The~CPMG sequence and CX implementation of the QAOA are employed throughout this~analysis.

\begin{figure*}[tb]
\includegraphics[width=0.48\linewidth]{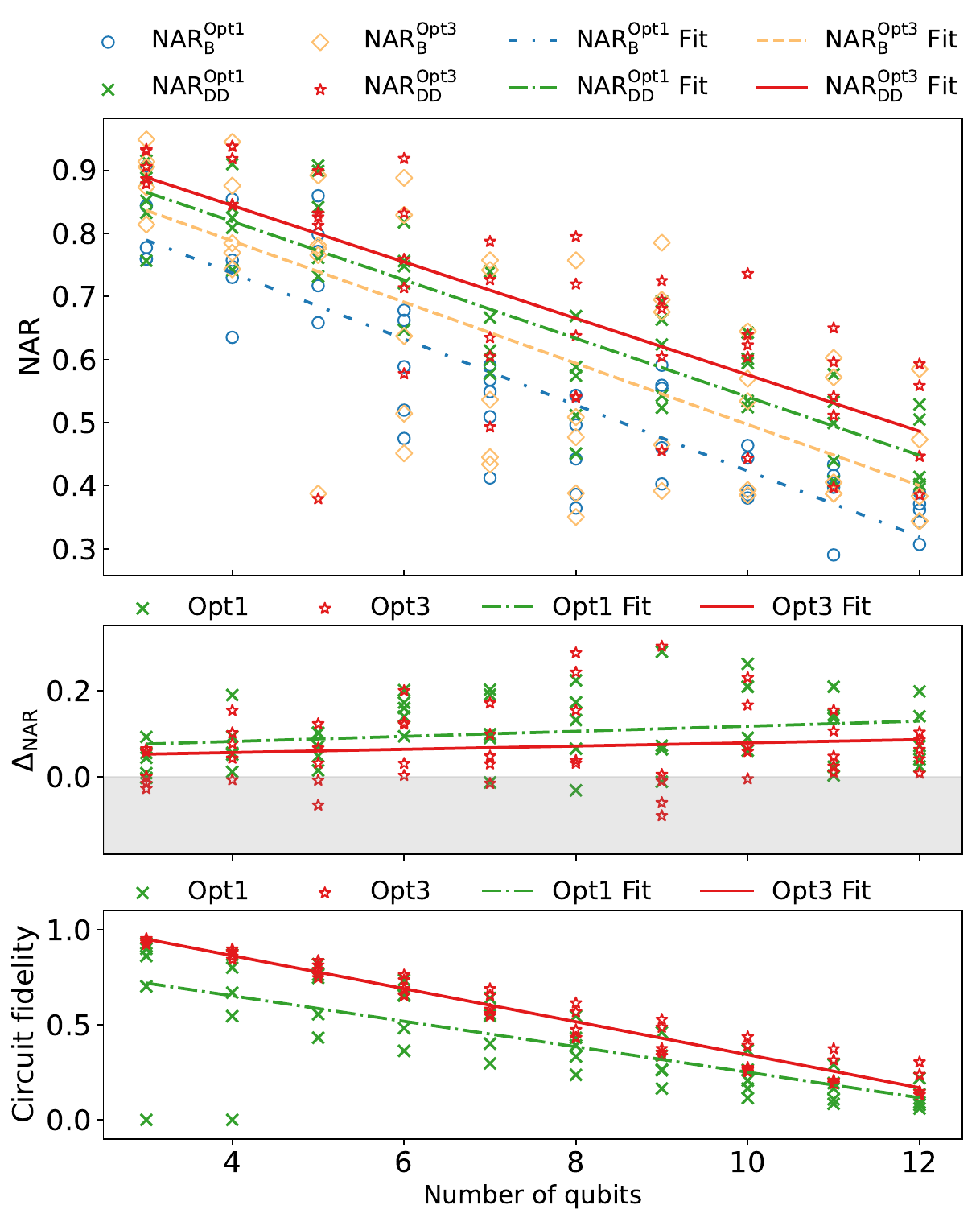}
\put(-0.48\linewidth,0.58\linewidth){(\textbf{a})}
\put(-0.48\linewidth,0.3\linewidth){(\textbf{c})}
\put(-0.48\linewidth,0.16\linewidth){(\textbf{e})}
\includegraphics[width=0.48\linewidth]{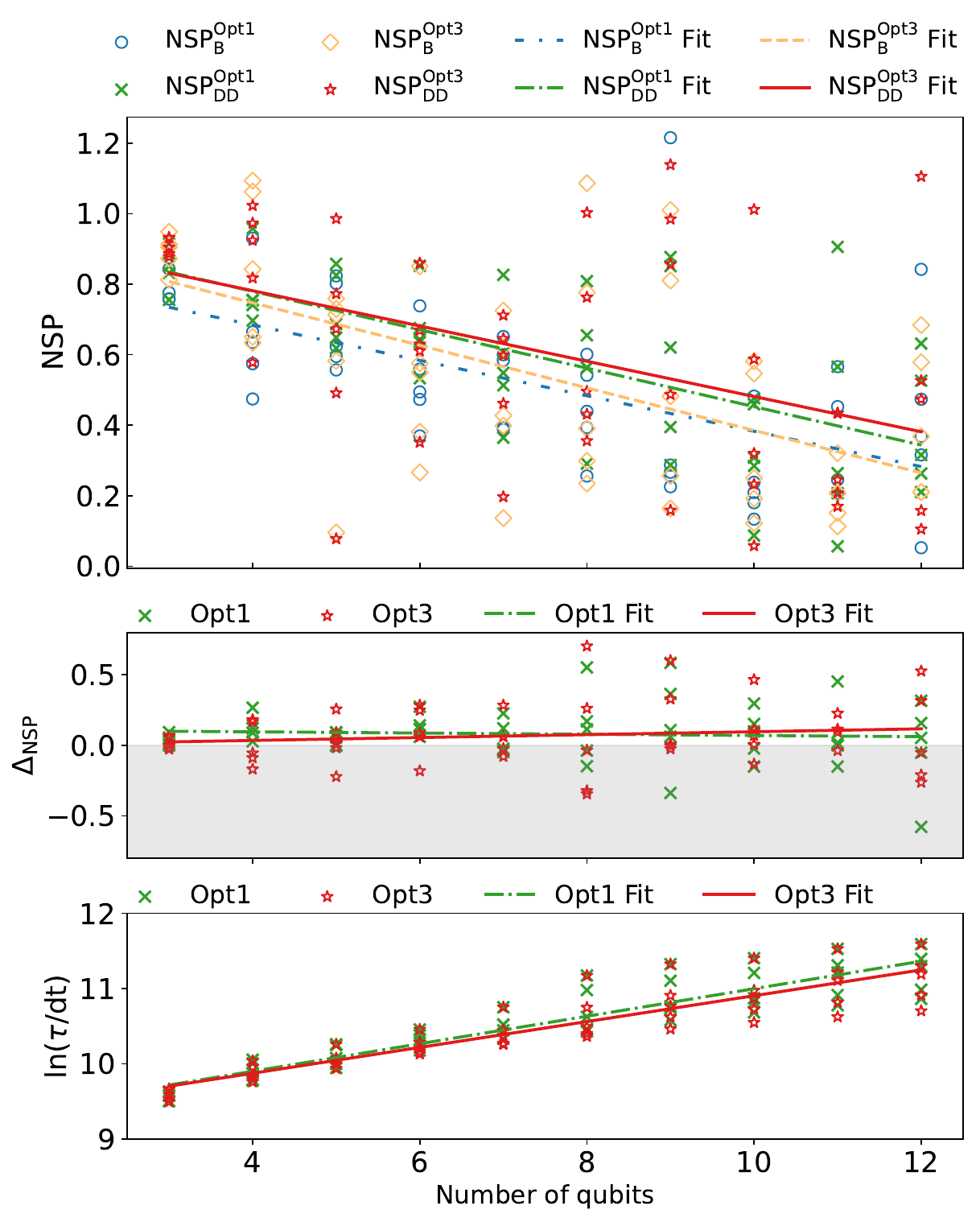}
\put(-0.48\linewidth,0.58\linewidth){(\textbf{b})}
\put(-0.48\linewidth,0.3\linewidth){(\textbf{d})}
\put(-0.48\linewidth,0.16\linewidth){(\textbf{f})}
\caption{\label{fig:optimization_level}
Comparison of optimization level 1 (Opt1) and optimization level 3 (Opt3) across five IBM QPUs ibmq\_kolkata, ibm\_cairo, ibmq\_ehningen, ibm\_cusco, and~ibm\_kyoto: (\textbf{a}) NAR, (\textbf{b}) NSP, (\textbf{c}) $\Delta_\mathrm{\scriptscriptstyle NAR}$, (\textbf{d}) $\Delta_\mathrm{\scriptscriptstyle NSP}$, (\textbf{e}) circuit fidelity, and~(\textbf{f}) {$\ln(\tau/\mathrm{dt})$}. The~CPMG sequence is used for all data points.
Each line represents a linear fit of the data.
}
\end{figure*}

\begin{table*}[tb]
    \caption{\label{tab:optimization_level} Parameters derived from the analysis of Fig.~\ref{fig:optimization_level}.}
    \centering
    \begin{ruledtabular}
    
\begin{tabular}{lllll}
 \textbf{Metric} & \textbf{Mean} & \textbf{Fit Function} & \textbf{Correlation Coefficient} & \textbf{\emph{p}-Value}\\
 \hline
 $\mathrm{NAR}_\mathrm{\scriptscriptstyle B}^\mathrm{\scriptscriptstyle Opt1}$ & 0.554 & $y=-0.052x+0.946$ &  $-0.891$ &  0 \\
 $\mathrm{NAR}_\mathrm{\scriptscriptstyle DD}^\mathrm{\scriptscriptstyle Opt1}$ & 0.657 & $y=-0.046x+1.004$ & $-0.886$  & 0 \\
 $\mathrm{NAR}_\mathrm{\scriptscriptstyle B}^\mathrm{\scriptscriptstyle Opt3}$ & {0.618} & {$y=-0.048+0.982$} &  {$-0.720$} &  0 \\
 $\mathrm{NAR}_\mathrm{\scriptscriptstyle DD}^\mathrm{\scriptscriptstyle Opt3}$ & {0.688} & {$y=-0.045x+1.023$} & {$-0.770$}  & 0 \\
 \hline
 $\mathrm{NSP}_\mathrm{\scriptscriptstyle B}^\mathrm{\scriptscriptstyle Opt1}$ & {0.509} & {$y=-0.050x+0.885$} &  {$-0.596$} &  0 \\
 $\mathrm{NSP}_\mathrm{\scriptscriptstyle DD}^\mathrm{\scriptscriptstyle Opt1}$ & {0.589} & {$y=-0.055x+0.998$} & {$-0.667$}  & {0} \\
 $\mathrm{NSP}_\mathrm{\scriptscriptstyle B}^\mathrm{\scriptscriptstyle Opt3}$ & 0.536 & $y=-0.060x+0.988$ &  $-0.586$ &  0.00001\\
 $\mathrm{NSP}_\mathrm{\scriptscriptstyle DD}^\mathrm{\scriptscriptstyle Opt3}$ & 0.606 & $y=-0.050x+0.981$ & $-0.471$  & 0.00056 \\
 
 \hline
 $\Delta_\mathrm{\scriptscriptstyle NAR}^\mathrm{\scriptscriptstyle Opt1}$ & 0.103 & $y=\phantom{-}0.006x+0.058$ &  $\phantom{-}0.218$ &  0.12785  \\
 $\Delta_\mathrm{\scriptscriptstyle NAR}^\mathrm{\scriptscriptstyle Opt3}$ & 0.069 & $y=\phantom{-}0.004x+0.041$ &  $\phantom{-}0.126$ & 0.38513 \\
 $\Delta_\mathrm{\scriptscriptstyle NSP}^\mathrm{\scriptscriptstyle Opt1}$ & 0.08 & $y=-0.004x+0.113$ &  $-0.065$ &  0.65396  \\
 $\Delta_\mathrm{\scriptscriptstyle NSP}^\mathrm{\scriptscriptstyle Opt3}$ & 0.07 & $y=\phantom{-}0.010x-0.007$ &  $\phantom{-}0.136$ & 0.34466  \\
 \hline
 Circuit fidelity (Opt1) & 0.417 & $y=-0.067x+0.919$ &  $-0.729$ & 0  \\
 Circuit fidelity (Opt3) & 0.559 & $y=-0.087x+1.209$ &  $-0.973$ & 0  \\
 {$\ln(\tau/\mathrm{dt})$} (Opt1) & 10.540 & $y=\phantom{-}0.183x+9.168$ &  $\phantom{-}0.919$ & 0  \\
 {$\ln(\tau/\mathrm{dt})$} (Opt3) & 10.475 & $y=\phantom{-}0.172x+9.186$ &  $\phantom{-}0.901$ & 0  \\
\end{tabular}
    \end{ruledtabular}

\end{table*}

Figures~\ref{fig:optimization_level}(a) and (b) demonstrate that Opt3 with DD sequences achieves the highest overall performance in terms of NAR and NSP{,} followed by Opt1 with DD sequences. Without~{error mitigation}, Opt3 outperforms Opt1 for all tested qubit counts in terms of NAR{, whereas}, for NSP, Opt3 exhibits an advantage only for a small number of qubits, with~comparable performance achieved at larger qubit counts.
{T}he average improvement in NAR due to DD ($\Delta_\mathrm{\scriptscriptstyle NAR}$) is generally higher for Opt1 compared to Opt3 {(Fig.~\ref{fig:optimization_level}(c))}. {However, for~NSP, DD initially benefits Opt1 more, but~this advantage shifts toward Opt3 for larger qubit counts (Fig.~\ref{fig:optimization_level}(d))}.
Furthermore, the~reported average $\mathrm{EMSR}_\mathrm{\scriptscriptstyle AR}$ and $\mathrm{EMSR}_\mathrm{\scriptscriptstyle SP}$ are 92\% and 74\% for Opt1, respectively, compared to 80\% and 60\% for Opt3{, suggesting a higher robustness of DD} for Opt1. It is important to note that Opt3 exhibits higher circuit fidelity (Fig.~\ref{fig:optimization_level}(e)) and a shorter schedule duration (Fig.~\ref{fig:optimization_level}(f)), which are crucial for high algorithm~performance.

The detailed linear fit parameters for the optimization levels are provided in Table~\ref{tab:optimization_level}. {While DD effectively mitigates the decrease in NAR for both Opt1 and Opt3 with increasing qubits, its impact on NSP differs. For~Opt3, DD suppresses the decrease in NSP, whereas, for Opt1, it appears to exacerbate the decay. The~correlation between NSP and qubit count for Opt1 is very weak (low $C_r$ and high $p$-value), suggesting minimal influence from qubit count on NSP for this optimization level. In~contrast, Opt1 prioritizes schedule duration, exhibiting a stronger correlation with qubit count, while Opt3 prioritizes circuit fidelity, showing a stronger correlation between fidelity and qubit count.}

This analysis demonstrates a trade-off between the optimization level and DD effectiveness. While Opt3 offers superior overall performance with DD sequences, Opt1 exhibits higher DD effectiveness. Furthermore, DD sequences become increasingly beneficial for Opt3 at larger qubit counts for finding optimal~solutions.

\section{Discussion and~Conclusions\label{sec:discu_conclu}}

Our comprehensive study, conducted on eight IBM quantum devices, reveals that the application of dynamical decoupling (DD) sequences can significantly enhance the performance and robustness of algorithms on near-term quantum devices. However, the~effectiveness of DD sequences varies depending on hardware and algorithmic factors. A~key finding is the observed inverse relationship between DD effectiveness and the original performance of algorithms without error mitigation. This implies that algorithms with higher inherent performance (measured without DD sequences) exhibit lower DD effectiveness. For~instance, ECR-based QPUs offer superior native performance but reduced DD effectiveness compared to CX-based QPUs. Similarly, the CZ implementation of a QAOA exhibits higher native algorithm performance but lower average DD effectiveness compared to CX implementation. Moreover, optimization level 3 produces higher algorithm performance, but~level 1 exhibits higher DD effectiveness. This inverse behavior can be attributed to the fact that algorithms with high performance typically have a lower intrinsic error rate, including the decoherence errors targeted by DD sequences. Furthermore, the~introduction of DD pulse sequences can lead to new gate operation errors that decrease algorithm performance and potentially limit their effectiveness for certain~algorithms.

DD sequences are typically more effective for algorithms with lower fidelity and a longer schedule duration, but~their impact on approximation ratio and success probability differs. The~results indicate that while algorithms with lower circuit fidelity struggle to achieve high approximation ratio values, the~application of DD sequences allows for achieving the simulated success probability. This finding suggests a potential methodology for obtaining the desired probabilities by studying different algorithm implementations and DD sequences and~selecting the measure that minimizes system~energy.

{Table~\ref{tab:summary_data} summarizes the observed effects of investigated hardware and algorithm factors on the effectiveness and robustness of DD. Without~error mitigation, the~ECR native gate set achieves the highest average value of normalized approximation ratio (NAR$_\text{B}$), while the CZ implementation of the QAOA exhibits the highest normalized success probability (NSP$_\text{B}$). However, when applying DD error mitigation, the~CX native gate set and CX implementation of the QAOA demonstrate the greatest increase in NAR, while the XY4 sequence leads to the most significant NSP improvement. Additionally, the~linear fit slope coefficients suggest that circuit fidelity has a stronger influence on approximation ratio, whereas schedule duration more significantly impacts success probability. Furthermore, as~the qubit count increases, the~CX gate set, CX implementation, and~XY4 sequence benefit more from DD mitigation compared to the ECR gate set, CZ implementation, and~CPMG sequence. Notably, the~CX implementation exhibits the highest overall robustness for the DD strategy, followed by the CX gate set and circuit optimization level 1 (Opt1).}

\begin{table*}
    \caption{\label{tab:summary_data}
Impact of hardware and algorithm factors on DD effectiveness and robustness.}
    \centering
    \begin{ruledtabular}
    
\begin{tabular}{llllllllll}
  \multirow[l]{2}{*}{\textbf{Factor}} & \multicolumn{4}{c}{\textbf{Mean}} & \multicolumn{2}{c}{\textbf{Slope Coefficient}} & \multicolumn{2}{c}{\textbf{EMSR}} \\
 & \boldmath{$\mathrm{NAR}_\mathrm{\scriptscriptstyle B}$} & \boldmath{$\Delta_\mathrm{\scriptscriptstyle NAR}$} & \boldmath{$\mathrm{NSP}_\mathrm{\scriptscriptstyle B}$} & \boldmath{$\Delta_\mathrm{\scriptscriptstyle NSP}$} & \boldmath{$\phantom{-}\Delta_\mathrm{\scriptscriptstyle NAR}$} & \boldmath{$\phantom{-}\Delta_\mathrm{\scriptscriptstyle NSP}$} & \boldmath{$\mathrm{EMSR}_\mathrm{\scriptscriptstyle AR}$} & \boldmath{$\mathrm{EMSR}_\mathrm{\scriptscriptstyle SP}$}\\
 \hline
 \multicolumn{1}{l}{Circuit fidelity} & \multirow[l]{2}{*}{0.610} & \multirow[l]{2}{*}{0.077} & \multirow[l]{2}{*}{0.556} & \multirow[l]{2}{*}{0.075} & $-0.065$ & $-0.021$ & \multirow[l]{2}{*}{85.55\%} & \multirow[l]{2}{*}{66.80\%}\\
 \multicolumn{1}{l}{Schedule duration} & & & & & $\phantom{-}0.042$ & $\phantom{-}0.033$ & \\
 \multirow[l]{1}{*}{CX gate set}  & 0.548 & $0.108$ & 0.464 & 0.118 & $\phantom{-}0.009$ & $\phantom{-}0.017$ & $92.50\%$ & $72.50\%$ \\
 \multirow[l]{1}{*}{ECR gate set} & $0.712$ & 0.059 & 0.618 & 0.098 & $\phantom{-}0.005$ & $\phantom{-}0.009$ & $75.00\%$ & $62.50\%$ \\
 \multirow[l]{1}{*}{CX implementation} & 0.548 & $0.108$ & 0.464 & 0.118 & $\phantom{-}0.009$ & $\phantom{-}0.017$ & $92.50\%$ & $80.00\%$\\
 \multirow[l]{1}{*}{CZ implementation} & 0.638 & 0.053 & $0.635$ & 0.018 & $\phantom{-}0.003$ & $-0.015$ & $75.00\%$ & $57.50\%$\\
 \multirow[l]{1}{*}{CPMG sequence} & 0.610 & 0.090 & 0.525 & 0.104 & $\phantom{-}0.006$ & $\phantom{-}0.011$ & $84.29\%$ & $75.71\%$ \\
 \multirow[l]{1}{*}{XY4 sequence} & 0.600 & 0.097 & 0.509 & $0.139$ & $\phantom{-}0.009$ & $\phantom{-}0.014$ & $67.14\%$ & $64.29\%$ \\
 \multirow[l]{1}{*}{Opt1} & 0.554 & 0.103 & 0.509 & 0.080 & $\phantom{-}0.006$ & $-0.004$ & $92.00\%$ & $74.00\%$ \\
 \multirow[l]{1}{*}{Opt3} & 0.618 & 0.069 & 0.536 & 0.070 & $\phantom{-}0.004$ & $\phantom{-}0.010$ & $80.00\%$ & $60.00\%$ \\
\end{tabular}
    \end{ruledtabular}

\end{table*}

One significant reason for the higher performance of algorithms on ECR-based QPUs is the inherently higher circuit fidelity and shorter schedule duration. However, another potential factor could be the inherent advantages of ECR gates. In~these devices, ECR gates are only allowed for one direction, meaning that any two-qubit gate is directly decomposed into ECR gates with the same direction on the qubit pair. While the hardware-native CX gate also has a native direction, the~reverse direction is supported, requiring additional single-qubit gates and the hardware-native CX gate for implementation. Directly decomposing algorithms into one-directional ECR gates could be more advantageous than using CX gates with bidirectional capability at the gate level and then transforming them into native CX gates at the hardware pulse level. A~similar trend is observed for CZ implementation, which produces a higher algorithm performance than CX implementation. The~CX gate is directed, whereas the CZ gate is undirected, allowing for the decomposition of all two-qubit gates with one directed CZ gate on one or more qubit pairs. Utilizing gates in the same direction could lead to a higher symmetry in the circuit, both in terms of single- and two-qubit gates, potentially contributing to error suppression and improved algorithmic performance. This highlights the importance of considering native {gate properties} at the pulse level and maintaining circuit structure symmetry during algorithm design. {Our findings hold broad applicability across various quantum algorithms, including the variational quantum eigensolver (VQE) \cite{peruzzo2014variational, tilly2022variational}. Prioritizing native gates and maintaining circuit symmetry during VQE and other quantum algorithm executions can enhance performance and mitigate errors across diverse quantum computing platforms.}

In conclusion, this study demonstrates the significant impact of several factors on algorithm performance and the effectiveness of error mitigation across eight IBM QPUs. These factors include circuit fidelity, schedule duration, choice of hardware-native gates, algorithm implementations, types of DD sequences, and~optimization levels. Despite minimal performance variations between Carr--Purcell--Meiboom--Gill (CPMG) and XY4 sequences, XY4 exhibits a slight advantage in success probability for larger qubit counts. However, CPMG achieves a higher overall error mitigation success rate, suggesting potentially greater robustness. 
While the results highlight the general enhancement of algorithm performance and robustness through the use of DD sequences, achieving significantly improved performance relies more critically on factors such as high-quality native gates of QPUs, symmetric algorithm implementation, and~effective circuit optimization techniques. Therefore, a~holistic approach that considers both hardware characteristics and software optimization strategies is important when designing quantum algorithms to maximize reliability and efficacy.
This study underscores the importance of hardware considerations and circuit design in enhancing algorithm performance and robustness using DD sequences. These insights guide the development and optimization of other quantum applications.
Future research directions include investigating the interplay between these factors when combining DD sequences with other error mitigation strategies, such as zero-noise extrapolation~\cite{li2017efficient,temme2017error}. Additionally, exploring the relationship between circuit symmetry and its effectiveness in suppressing errors could provide valuable insights. Extending this analysis to other quantum algorithms, such as the {VQE} and protocols for preparing Greenberger--Horne--Zeilinger (GHZ) states, holds promise for revealing the broader applicability of these~findings.
\\

\begin{acknowledgments}

We acknowledge the use of IBM Quantum services for this work and to advanced services provided by the IBM Quantum Researchers Program. The views expressed are those of the authors, and do not reflect the official policy or position of IBM or the IBM Quantum team.
This work was funded in part by the Ministry of Economic Affairs, Labour and Tourism Baden Württemberg, under the project QORA in the frame of the Competence Center Quantum Computing Baden-Württemberg.

\end{acknowledgments}

\appendix


\nocite{*}

\bibliography{refs}

\begin{thebibliography}{55}%
\makeatletter
\providecommand \@ifxundefined [1]{%
 \@ifx{#1\undefined}
}%
\providecommand \@ifnum [1]{%
 \ifnum #1\expandafter \@firstoftwo
 \else \expandafter \@secondoftwo
 \fi
}%
\providecommand \@ifx [1]{%
 \ifx #1\expandafter \@firstoftwo
 \else \expandafter \@secondoftwo
 \fi
}%
\providecommand \natexlab [1]{#1}%
\providecommand \enquote  [1]{``#1''}%
\providecommand \bibnamefont  [1]{#1}%
\providecommand \bibfnamefont [1]{#1}%
\providecommand \citenamefont [1]{#1}%
\providecommand \href@noop [0]{\@secondoftwo}%
\providecommand \href [0]{\begingroup \@sanitize@url \@href}%
\providecommand \@href[1]{\@@startlink{#1}\@@href}%
\providecommand \@@href[1]{\endgroup#1\@@endlink}%
\providecommand \@sanitize@url [0]{\catcode `\\12\catcode `\$12\catcode `\&12\catcode `\#12\catcode `\^12\catcode `\_12\catcode `\%12\relax}%
\providecommand \@@startlink[1]{}%
\providecommand \@@endlink[0]{}%
\providecommand \url  [0]{\begingroup\@sanitize@url \@url }%
\providecommand \@url [1]{\endgroup\@href {#1}{\urlprefix }}%
\providecommand \urlprefix  [0]{URL }%
\providecommand \Eprint [0]{\href }%
\providecommand \doibase [0]{https://doi.org/}%
\providecommand \selectlanguage [0]{\@gobble}%
\providecommand \bibinfo  [0]{\@secondoftwo}%
\providecommand \bibfield  [0]{\@secondoftwo}%
\providecommand \translation [1]{[#1]}%
\providecommand \BibitemOpen [0]{}%
\providecommand \bibitemStop [0]{}%
\providecommand \bibitemNoStop [0]{.\EOS\space}%
\providecommand \EOS [0]{\spacefactor3000\relax}%
\providecommand \BibitemShut  [1]{\csname bibitem#1\endcsname}%
\let\auto@bib@innerbib\@empty
\bibitem [{\citenamefont {Preskill}(2018)}]{preskill2018quantum}%
  \BibitemOpen
  \bibfield  {author} {\bibinfo {author} {\bibfnamefont {J.}~\bibnamefont {Preskill}},\ }\bibfield  {title} {\bibinfo {title} {Quantum computing in the nisq era and beyond},\ }\href {https://doi.org/10.22331/q-2018-08-06-79} {\bibfield  {journal} {\bibinfo  {journal} {Quantum}\ }\textbf {\bibinfo {volume} {2}},\ \bibinfo {pages} {79} (\bibinfo {year} {2018})}\BibitemShut {NoStop}%
\bibitem [{\citenamefont {Cai}\ \emph {et~al.}(2023)\citenamefont {Cai}, \citenamefont {Babbush}, \citenamefont {Benjamin}, \citenamefont {Endo}, \citenamefont {Huggins}, \citenamefont {Li}, \citenamefont {McClean},\ and\ \citenamefont {O'Brien}}]{cai2023quantum}%
  \BibitemOpen
  \bibfield  {author} {\bibinfo {author} {\bibfnamefont {Z.}~\bibnamefont {Cai}}, \bibinfo {author} {\bibfnamefont {R.}~\bibnamefont {Babbush}}, \bibinfo {author} {\bibfnamefont {S.~C.}\ \bibnamefont {Benjamin}}, \bibinfo {author} {\bibfnamefont {S.}~\bibnamefont {Endo}}, \bibinfo {author} {\bibfnamefont {W.~J.}\ \bibnamefont {Huggins}}, \bibinfo {author} {\bibfnamefont {Y.}~\bibnamefont {Li}}, \bibinfo {author} {\bibfnamefont {J.~R.}\ \bibnamefont {McClean}},\ and\ \bibinfo {author} {\bibfnamefont {T.~E.}\ \bibnamefont {O'Brien}},\ }\bibfield  {title} {\bibinfo {title} {Quantum error mitigation},\ }\href {https://doi.org/10.1103/RevModPhys.95.045005} {\bibfield  {journal} {\bibinfo  {journal} {Rev. Mod. Phys.}\ }\textbf {\bibinfo {volume} {95}},\ \bibinfo {pages} {045005} (\bibinfo {year} {2023})}\BibitemShut {NoStop}%
\bibitem [{\citenamefont {Suter}\ and\ \citenamefont {\'Alvarez}(2016)}]{suter2016colloquium}%
  \BibitemOpen
  \bibfield  {author} {\bibinfo {author} {\bibfnamefont {D.}~\bibnamefont {Suter}}\ and\ \bibinfo {author} {\bibfnamefont {G.~A.}\ \bibnamefont {\'Alvarez}},\ }\bibfield  {title} {\bibinfo {title} {Colloquium: Protecting quantum information against environmental noise},\ }\href {https://doi.org/10.1103/RevModPhys.88.041001} {\bibfield  {journal} {\bibinfo  {journal} {Rev. Mod. Phys.}\ }\textbf {\bibinfo {volume} {88}},\ \bibinfo {pages} {041001} (\bibinfo {year} {2016})}\BibitemShut {NoStop}%
\bibitem [{\citenamefont {Ali~Ahmed}\ \emph {et~al.}(2013)\citenamefont {Ali~Ahmed}, \citenamefont {\'Alvarez},\ and\ \citenamefont {Suter}}]{ahmed2013robustness}%
  \BibitemOpen
  \bibfield  {author} {\bibinfo {author} {\bibfnamefont {M.~A.}\ \bibnamefont {Ali~Ahmed}}, \bibinfo {author} {\bibfnamefont {G.~A.}\ \bibnamefont {\'Alvarez}},\ and\ \bibinfo {author} {\bibfnamefont {D.}~\bibnamefont {Suter}},\ }\bibfield  {title} {\bibinfo {title} {Robustness of dynamical decoupling sequences},\ }\href {https://doi.org/10.1103/PhysRevA.87.042309} {\bibfield  {journal} {\bibinfo  {journal} {Phys. Rev. A}\ }\textbf {\bibinfo {volume} {87}},\ \bibinfo {pages} {042309} (\bibinfo {year} {2013})}\BibitemShut {NoStop}%
\bibitem [{\citenamefont {Pokharel}\ \emph {et~al.}(2018)\citenamefont {Pokharel}, \citenamefont {Anand}, \citenamefont {Fortman},\ and\ \citenamefont {Lidar}}]{pokharel2018demonstration}%
  \BibitemOpen
  \bibfield  {author} {\bibinfo {author} {\bibfnamefont {B.}~\bibnamefont {Pokharel}}, \bibinfo {author} {\bibfnamefont {N.}~\bibnamefont {Anand}}, \bibinfo {author} {\bibfnamefont {B.}~\bibnamefont {Fortman}},\ and\ \bibinfo {author} {\bibfnamefont {D.~A.}\ \bibnamefont {Lidar}},\ }\bibfield  {title} {\bibinfo {title} {Demonstration of fidelity improvement using dynamical decoupling with superconducting qubits},\ }\href {https://doi.org/10.1103/PhysRevLett.121.220502} {\bibfield  {journal} {\bibinfo  {journal} {Phys. Rev. Lett.}\ }\textbf {\bibinfo {volume} {121}},\ \bibinfo {pages} {220502} (\bibinfo {year} {2018})}\BibitemShut {NoStop}%
\bibitem [{\citenamefont {Souza}\ \emph {et~al.}(2012{\natexlab{a}})\citenamefont {Souza}, \citenamefont {{\'A}lvarez},\ and\ \citenamefont {Suter}}]{souza2012robust}%
  \BibitemOpen
  \bibfield  {author} {\bibinfo {author} {\bibfnamefont {A.~M.}\ \bibnamefont {Souza}}, \bibinfo {author} {\bibfnamefont {G.~A.}\ \bibnamefont {{\'A}lvarez}},\ and\ \bibinfo {author} {\bibfnamefont {D.}~\bibnamefont {Suter}},\ }\bibfield  {title} {\bibinfo {title} {Robust dynamical decoupling},\ }\href@noop {} {\bibfield  {journal} {\bibinfo  {journal} {Philosophical Transactions of the Royal Society A: Mathematical, Physical and Engineering Sciences}\ }\textbf {\bibinfo {volume} {370}},\ \bibinfo {pages} {4748} (\bibinfo {year} {2012}{\natexlab{a}})}\BibitemShut {NoStop}%
\bibitem [{\citenamefont {De~Lange}\ \emph {et~al.}(2010)\citenamefont {De~Lange}, \citenamefont {Wang}, \citenamefont {Riste}, \citenamefont {Dobrovitski},\ and\ \citenamefont {Hanson}}]{de2010universal}%
  \BibitemOpen
  \bibfield  {author} {\bibinfo {author} {\bibfnamefont {G.}~\bibnamefont {De~Lange}}, \bibinfo {author} {\bibfnamefont {Z.}~\bibnamefont {Wang}}, \bibinfo {author} {\bibfnamefont {D.}~\bibnamefont {Riste}}, \bibinfo {author} {\bibfnamefont {V.}~\bibnamefont {Dobrovitski}},\ and\ \bibinfo {author} {\bibfnamefont {R.}~\bibnamefont {Hanson}},\ }\bibfield  {title} {\bibinfo {title} {Universal dynamical decoupling of a single solid-state spin from a spin bath},\ }\href@noop {} {\bibfield  {journal} {\bibinfo  {journal} {Science}\ }\textbf {\bibinfo {volume} {330}},\ \bibinfo {pages} {60} (\bibinfo {year} {2010})}\BibitemShut {NoStop}%
\bibitem [{\citenamefont {Du}\ \emph {et~al.}(2009)\citenamefont {Du}, \citenamefont {Rong}, \citenamefont {Zhao}, \citenamefont {Wang}, \citenamefont {Yang},\ and\ \citenamefont {Liu}}]{du2009preserving}%
  \BibitemOpen
  \bibfield  {author} {\bibinfo {author} {\bibfnamefont {J.}~\bibnamefont {Du}}, \bibinfo {author} {\bibfnamefont {X.}~\bibnamefont {Rong}}, \bibinfo {author} {\bibfnamefont {N.}~\bibnamefont {Zhao}}, \bibinfo {author} {\bibfnamefont {Y.}~\bibnamefont {Wang}}, \bibinfo {author} {\bibfnamefont {J.}~\bibnamefont {Yang}},\ and\ \bibinfo {author} {\bibfnamefont {R.}~\bibnamefont {Liu}},\ }\bibfield  {title} {\bibinfo {title} {Preserving electron spin coherence in solids by optimal dynamical decoupling},\ }\href@noop {} {\bibfield  {journal} {\bibinfo  {journal} {Nature}\ }\textbf {\bibinfo {volume} {461}},\ \bibinfo {pages} {1265} (\bibinfo {year} {2009})}\BibitemShut {NoStop}%
\bibitem [{\citenamefont {Farfurnik}\ \emph {et~al.}(2016)\citenamefont {Farfurnik}, \citenamefont {Jarmola}, \citenamefont {Pham}, \citenamefont {Wang}, \citenamefont {Dobrovitski}, \citenamefont {Walsworth}, \citenamefont {Budker},\ and\ \citenamefont {Bar-Gill}}]{farfurnik2016improving}%
  \BibitemOpen
  \bibfield  {author} {\bibinfo {author} {\bibfnamefont {D.}~\bibnamefont {Farfurnik}}, \bibinfo {author} {\bibfnamefont {A.}~\bibnamefont {Jarmola}}, \bibinfo {author} {\bibfnamefont {L.}~\bibnamefont {Pham}}, \bibinfo {author} {\bibfnamefont {Z.}~\bibnamefont {Wang}}, \bibinfo {author} {\bibfnamefont {V.}~\bibnamefont {Dobrovitski}}, \bibinfo {author} {\bibfnamefont {R.}~\bibnamefont {Walsworth}}, \bibinfo {author} {\bibfnamefont {D.}~\bibnamefont {Budker}},\ and\ \bibinfo {author} {\bibfnamefont {N.}~\bibnamefont {Bar-Gill}},\ }\bibfield  {title} {\bibinfo {title} {Improving the coherence properties of solid-state spin ensembles via optimized dynamical decoupling},\ }in\ \href@noop {} {\emph {\bibinfo {booktitle} {Quantum Optics}}},\ Vol.\ \bibinfo {volume} {9900}\ (\bibinfo {organization} {SPIE},\ \bibinfo {year} {2016})\ pp.\ \bibinfo {pages} {111--120}\BibitemShut {NoStop}%
\bibitem [{\citenamefont {Farfurnik}\ \emph {et~al.}(2015)\citenamefont {Farfurnik}, \citenamefont {Jarmola}, \citenamefont {Pham}, \citenamefont {Wang}, \citenamefont {Dobrovitski}, \citenamefont {Walsworth}, \citenamefont {Budker},\ and\ \citenamefont {Bar-Gill}}]{farfurnik2015optimizing}%
  \BibitemOpen
  \bibfield  {author} {\bibinfo {author} {\bibfnamefont {D.}~\bibnamefont {Farfurnik}}, \bibinfo {author} {\bibfnamefont {A.}~\bibnamefont {Jarmola}}, \bibinfo {author} {\bibfnamefont {L.~M.}\ \bibnamefont {Pham}}, \bibinfo {author} {\bibfnamefont {Z.-H.}\ \bibnamefont {Wang}}, \bibinfo {author} {\bibfnamefont {V.~V.}\ \bibnamefont {Dobrovitski}}, \bibinfo {author} {\bibfnamefont {R.~L.}\ \bibnamefont {Walsworth}}, \bibinfo {author} {\bibfnamefont {D.}~\bibnamefont {Budker}},\ and\ \bibinfo {author} {\bibfnamefont {N.}~\bibnamefont {Bar-Gill}},\ }\bibfield  {title} {\bibinfo {title} {Optimizing a dynamical decoupling protocol for solid-state electronic spin ensembles in diamond},\ }\href@noop {} {\bibfield  {journal} {\bibinfo  {journal} {Physical Review B}\ }\textbf {\bibinfo {volume} {92}},\ \bibinfo {pages} {060301} (\bibinfo {year} {2015})}\BibitemShut {NoStop}%
\bibitem [{\citenamefont {Merkel}\ \emph {et~al.}(2021)\citenamefont {Merkel}, \citenamefont {Cova Fari\~na},\ and\ \citenamefont {Reiserer}}]{merkel2021dynamical}%
  \BibitemOpen
  \bibfield  {author} {\bibinfo {author} {\bibfnamefont {B.}~\bibnamefont {Merkel}}, \bibinfo {author} {\bibfnamefont {P.}~\bibnamefont {Cova Fari\~na}},\ and\ \bibinfo {author} {\bibfnamefont {A.}~\bibnamefont {Reiserer}},\ }\bibfield  {title} {\bibinfo {title} {Dynamical decoupling of spin ensembles with strong anisotropic interactions},\ }\href {https://doi.org/10.1103/PhysRevLett.127.030501} {\bibfield  {journal} {\bibinfo  {journal} {Phys. Rev. Lett.}\ }\textbf {\bibinfo {volume} {127}},\ \bibinfo {pages} {030501} (\bibinfo {year} {2021})}\BibitemShut {NoStop}%
\bibitem [{\citenamefont {Medford}\ \emph {et~al.}(2012)\citenamefont {Medford}, \citenamefont {Barthel}, \citenamefont {Marcus}, \citenamefont {Hanson}, \citenamefont {Gossard} \emph {et~al.}}]{medford2012scaling}%
  \BibitemOpen
  \bibfield  {author} {\bibinfo {author} {\bibfnamefont {J.}~\bibnamefont {Medford}}, \bibinfo {author} {\bibfnamefont {C.}~\bibnamefont {Barthel}}, \bibinfo {author} {\bibfnamefont {C.}~\bibnamefont {Marcus}}, \bibinfo {author} {\bibfnamefont {M.}~\bibnamefont {Hanson}}, \bibinfo {author} {\bibfnamefont {A.}~\bibnamefont {Gossard}}, \emph {et~al.},\ }\bibfield  {title} {\bibinfo {title} {Scaling of dynamical decoupling for spin qubits},\ }\href@noop {} {\bibfield  {journal} {\bibinfo  {journal} {Physical review letters}\ }\textbf {\bibinfo {volume} {108}},\ \bibinfo {pages} {086802} (\bibinfo {year} {2012})}\BibitemShut {NoStop}%
\bibitem [{\citenamefont {Tripathi}\ \emph {et~al.}(2022)\citenamefont {Tripathi}, \citenamefont {Chen}, \citenamefont {Khezri}, \citenamefont {Yip}, \citenamefont {Levenson-Falk},\ and\ \citenamefont {Lidar}}]{tripathi2022suppression}%
  \BibitemOpen
  \bibfield  {author} {\bibinfo {author} {\bibfnamefont {V.}~\bibnamefont {Tripathi}}, \bibinfo {author} {\bibfnamefont {H.}~\bibnamefont {Chen}}, \bibinfo {author} {\bibfnamefont {M.}~\bibnamefont {Khezri}}, \bibinfo {author} {\bibfnamefont {K.-W.}\ \bibnamefont {Yip}}, \bibinfo {author} {\bibfnamefont {E.}~\bibnamefont {Levenson-Falk}},\ and\ \bibinfo {author} {\bibfnamefont {D.~A.}\ \bibnamefont {Lidar}},\ }\bibfield  {title} {\bibinfo {title} {Suppression of crosstalk in superconducting qubits using dynamical decoupling},\ }\href@noop {} {\bibfield  {journal} {\bibinfo  {journal} {Physical Review Applied}\ }\textbf {\bibinfo {volume} {18}},\ \bibinfo {pages} {024068} (\bibinfo {year} {2022})}\BibitemShut {NoStop}%
\bibitem [{\citenamefont {Bylander}\ \emph {et~al.}(2011)\citenamefont {Bylander}, \citenamefont {Gustavsson}, \citenamefont {Yan}, \citenamefont {Yoshihara}, \citenamefont {Harrabi}, \citenamefont {Fitch}, \citenamefont {Cory}, \citenamefont {Nakamura}, \citenamefont {Tsai},\ and\ \citenamefont {Oliver}}]{bylander2011noise}%
  \BibitemOpen
  \bibfield  {author} {\bibinfo {author} {\bibfnamefont {J.}~\bibnamefont {Bylander}}, \bibinfo {author} {\bibfnamefont {S.}~\bibnamefont {Gustavsson}}, \bibinfo {author} {\bibfnamefont {F.}~\bibnamefont {Yan}}, \bibinfo {author} {\bibfnamefont {F.}~\bibnamefont {Yoshihara}}, \bibinfo {author} {\bibfnamefont {K.}~\bibnamefont {Harrabi}}, \bibinfo {author} {\bibfnamefont {G.}~\bibnamefont {Fitch}}, \bibinfo {author} {\bibfnamefont {D.~G.}\ \bibnamefont {Cory}}, \bibinfo {author} {\bibfnamefont {Y.}~\bibnamefont {Nakamura}}, \bibinfo {author} {\bibfnamefont {J.-S.}\ \bibnamefont {Tsai}},\ and\ \bibinfo {author} {\bibfnamefont {W.~D.}\ \bibnamefont {Oliver}},\ }\bibfield  {title} {\bibinfo {title} {Noise spectroscopy through dynamical decoupling with a superconducting flux qubit},\ }\href@noop {} {\bibfield  {journal} {\bibinfo  {journal} {Nature Physics}\ }\textbf {\bibinfo {volume} {7}},\ \bibinfo {pages} {565} (\bibinfo {year} {2011})}\BibitemShut {NoStop}%
\bibitem [{\citenamefont {Biercuk}\ \emph {et~al.}(2009)\citenamefont {Biercuk}, \citenamefont {Uys}, \citenamefont {VanDevender}, \citenamefont {Shiga}, \citenamefont {Itano},\ and\ \citenamefont {Bollinger}}]{biercuk2009experimental}%
  \BibitemOpen
  \bibfield  {author} {\bibinfo {author} {\bibfnamefont {M.~J.}\ \bibnamefont {Biercuk}}, \bibinfo {author} {\bibfnamefont {H.}~\bibnamefont {Uys}}, \bibinfo {author} {\bibfnamefont {A.~P.}\ \bibnamefont {VanDevender}}, \bibinfo {author} {\bibfnamefont {N.}~\bibnamefont {Shiga}}, \bibinfo {author} {\bibfnamefont {W.~M.}\ \bibnamefont {Itano}},\ and\ \bibinfo {author} {\bibfnamefont {J.~J.}\ \bibnamefont {Bollinger}},\ }\bibfield  {title} {\bibinfo {title} {Experimental uhrig dynamical decoupling using trapped ions},\ }\href {https://doi.org/10.1103/PhysRevA.79.062324} {\bibfield  {journal} {\bibinfo  {journal} {Phys. Rev. A}\ }\textbf {\bibinfo {volume} {79}},\ \bibinfo {pages} {062324} (\bibinfo {year} {2009})}\BibitemShut {NoStop}%
\bibitem [{\citenamefont {Evert}\ \emph {et~al.}(2024)\citenamefont {Evert}, \citenamefont {Izquierdo}, \citenamefont {Sud}, \citenamefont {Hu}, \citenamefont {Grabbe}, \citenamefont {Rieffel}, \citenamefont {Reagor},\ and\ \citenamefont {Wang}}]{evert2024syncopated}%
  \BibitemOpen
  \bibfield  {author} {\bibinfo {author} {\bibfnamefont {B.}~\bibnamefont {Evert}}, \bibinfo {author} {\bibfnamefont {Z.~G.}\ \bibnamefont {Izquierdo}}, \bibinfo {author} {\bibfnamefont {J.}~\bibnamefont {Sud}}, \bibinfo {author} {\bibfnamefont {H.-Y.}\ \bibnamefont {Hu}}, \bibinfo {author} {\bibfnamefont {S.}~\bibnamefont {Grabbe}}, \bibinfo {author} {\bibfnamefont {E.~G.}\ \bibnamefont {Rieffel}}, \bibinfo {author} {\bibfnamefont {M.~J.}\ \bibnamefont {Reagor}},\ and\ \bibinfo {author} {\bibfnamefont {Z.}~\bibnamefont {Wang}},\ }\bibfield  {title} {\bibinfo {title} {Syncopated dynamical decoupling for suppressing crosstalk in quantum circuits},\ }\href@noop {} {\bibfield  {journal} {\bibinfo  {journal} {arXiv preprint arXiv:2403.07836}\ } (\bibinfo {year} {2024})}\BibitemShut {NoStop}%
\bibitem [{\citenamefont {Zhou}\ \emph {et~al.}(2023)\citenamefont {Zhou}, \citenamefont {Sitler}, \citenamefont {Oda}, \citenamefont {Schultz},\ and\ \citenamefont {Quiroz}}]{zhou2023quantum}%
  \BibitemOpen
  \bibfield  {author} {\bibinfo {author} {\bibfnamefont {Z.}~\bibnamefont {Zhou}}, \bibinfo {author} {\bibfnamefont {R.}~\bibnamefont {Sitler}}, \bibinfo {author} {\bibfnamefont {Y.}~\bibnamefont {Oda}}, \bibinfo {author} {\bibfnamefont {K.}~\bibnamefont {Schultz}},\ and\ \bibinfo {author} {\bibfnamefont {G.}~\bibnamefont {Quiroz}},\ }\bibfield  {title} {\bibinfo {title} {Quantum crosstalk robust quantum control},\ }\href {https://doi.org/10.1103/PhysRevLett.131.210802} {\bibfield  {journal} {\bibinfo  {journal} {Phys. Rev. Lett.}\ }\textbf {\bibinfo {volume} {131}},\ \bibinfo {pages} {210802} (\bibinfo {year} {2023})}\BibitemShut {NoStop}%
\bibitem [{\citenamefont {Shirizly}\ \emph {et~al.}(2024)\citenamefont {Shirizly}, \citenamefont {Misguich},\ and\ \citenamefont {Landa}}]{shirizly2024dissipative}%
  \BibitemOpen
  \bibfield  {author} {\bibinfo {author} {\bibfnamefont {L.}~\bibnamefont {Shirizly}}, \bibinfo {author} {\bibfnamefont {G.}~\bibnamefont {Misguich}},\ and\ \bibinfo {author} {\bibfnamefont {H.}~\bibnamefont {Landa}},\ }\bibfield  {title} {\bibinfo {title} {Dissipative dynamics of graph-state stabilizers with superconducting qubits},\ }\href {https://doi.org/10.1103/PhysRevLett.132.010601} {\bibfield  {journal} {\bibinfo  {journal} {Phys. Rev. Lett.}\ }\textbf {\bibinfo {volume} {132}},\ \bibinfo {pages} {010601} (\bibinfo {year} {2024})}\BibitemShut {NoStop}%
\bibitem [{\citenamefont {Seif}\ \emph {et~al.}(2024)\citenamefont {Seif}, \citenamefont {Liao}, \citenamefont {Tripathi}, \citenamefont {Krsulich}, \citenamefont {Malekakhlagh}, \citenamefont {Amico}, \citenamefont {Jurcevic},\ and\ \citenamefont {Javadi-Abhari}}]{seif2024suppressing}%
  \BibitemOpen
  \bibfield  {author} {\bibinfo {author} {\bibfnamefont {A.}~\bibnamefont {Seif}}, \bibinfo {author} {\bibfnamefont {H.}~\bibnamefont {Liao}}, \bibinfo {author} {\bibfnamefont {V.}~\bibnamefont {Tripathi}}, \bibinfo {author} {\bibfnamefont {K.}~\bibnamefont {Krsulich}}, \bibinfo {author} {\bibfnamefont {M.}~\bibnamefont {Malekakhlagh}}, \bibinfo {author} {\bibfnamefont {M.}~\bibnamefont {Amico}}, \bibinfo {author} {\bibfnamefont {P.}~\bibnamefont {Jurcevic}},\ and\ \bibinfo {author} {\bibfnamefont {A.}~\bibnamefont {Javadi-Abhari}},\ }\bibfield  {title} {\bibinfo {title} {Suppressing correlated noise in quantum computers via context-aware compiling},\ }\href@noop {} {\bibfield  {journal} {\bibinfo  {journal} {arXiv preprint arXiv:2403.06852}\ } (\bibinfo {year} {2024})}\BibitemShut {NoStop}%
\bibitem [{\citenamefont {Qiu}\ \emph {et~al.}(2021)\citenamefont {Qiu}, \citenamefont {Zhou}, \citenamefont {Hu}, \citenamefont {Yuan}, \citenamefont {Zhang}, \citenamefont {Chu}, \citenamefont {Huang}, \citenamefont {Liu}, \citenamefont {Luo}, \citenamefont {Ni} \emph {et~al.}}]{qiu2021suppressing}%
  \BibitemOpen
  \bibfield  {author} {\bibinfo {author} {\bibfnamefont {J.}~\bibnamefont {Qiu}}, \bibinfo {author} {\bibfnamefont {Y.}~\bibnamefont {Zhou}}, \bibinfo {author} {\bibfnamefont {C.-K.}\ \bibnamefont {Hu}}, \bibinfo {author} {\bibfnamefont {J.}~\bibnamefont {Yuan}}, \bibinfo {author} {\bibfnamefont {L.}~\bibnamefont {Zhang}}, \bibinfo {author} {\bibfnamefont {J.}~\bibnamefont {Chu}}, \bibinfo {author} {\bibfnamefont {W.}~\bibnamefont {Huang}}, \bibinfo {author} {\bibfnamefont {W.}~\bibnamefont {Liu}}, \bibinfo {author} {\bibfnamefont {K.}~\bibnamefont {Luo}}, \bibinfo {author} {\bibfnamefont {Z.}~\bibnamefont {Ni}}, \emph {et~al.},\ }\bibfield  {title} {\bibinfo {title} {Suppressing coherent two-qubit errors via dynamical decoupling},\ }\href@noop {} {\bibfield  {journal} {\bibinfo  {journal} {Physical Review Applied}\ }\textbf {\bibinfo {volume} {16}},\ \bibinfo {pages} {054047} (\bibinfo {year} {2021})}\BibitemShut {NoStop}%
\bibitem [{\citenamefont {Carr}\ and\ \citenamefont {Purcell}(1954)}]{carr1954effects}%
  \BibitemOpen
  \bibfield  {author} {\bibinfo {author} {\bibfnamefont {H.~Y.}\ \bibnamefont {Carr}}\ and\ \bibinfo {author} {\bibfnamefont {E.~M.}\ \bibnamefont {Purcell}},\ }\bibfield  {title} {\bibinfo {title} {Effects of diffusion on free precession in nuclear magnetic resonance experiments},\ }\href {https://doi.org/10.1103/PhysRev.94.630} {\bibfield  {journal} {\bibinfo  {journal} {Phys. Rev.}\ }\textbf {\bibinfo {volume} {94}},\ \bibinfo {pages} {630} (\bibinfo {year} {1954})}\BibitemShut {NoStop}%
\bibitem [{\citenamefont {Meiboom}\ and\ \citenamefont {Gill}(2004)}]{meiboom2004modified}%
  \BibitemOpen
  \bibfield  {author} {\bibinfo {author} {\bibfnamefont {S.}~\bibnamefont {Meiboom}}\ and\ \bibinfo {author} {\bibfnamefont {D.}~\bibnamefont {Gill}},\ }\bibfield  {title} {\bibinfo {title} {{Modified Spin‐Echo Method for Measuring Nuclear Relaxation Times}},\ }\href {https://doi.org/10.1063/1.1716296} {\bibfield  {journal} {\bibinfo  {journal} {Review of Scientific Instruments}\ }\textbf {\bibinfo {volume} {29}},\ \bibinfo {pages} {688} (\bibinfo {year} {2004})},\ \Eprint {https://arxiv.org/abs/https://pubs.aip.org/aip/rsi/article-pdf/29/8/688/8343239/688\_1\_online.pdf} {https://pubs.aip.org/aip/rsi/article-pdf/29/8/688/8343239/688\_1\_online.pdf} \BibitemShut {NoStop}%
\bibitem [{\citenamefont {Maudsley}(1986)}]{maudsley1986modified}%
  \BibitemOpen
  \bibfield  {author} {\bibinfo {author} {\bibfnamefont {A.}~\bibnamefont {Maudsley}},\ }\bibfield  {title} {\bibinfo {title} {Modified carr-purcell-meiboom-gill sequence for nmr fourier imaging applications},\ }\href {https://doi.org/https://doi.org/10.1016/0022-2364(86)90160-5} {\bibfield  {journal} {\bibinfo  {journal} {Journal of Magnetic Resonance (1969)}\ }\textbf {\bibinfo {volume} {69}},\ \bibinfo {pages} {488} (\bibinfo {year} {1986})}\BibitemShut {NoStop}%
\bibitem [{\citenamefont {\'Alvarez}\ \emph {et~al.}(2012)\citenamefont {\'Alvarez}, \citenamefont {Souza},\ and\ \citenamefont {Suter}}]{alvarez2012iterative}%
  \BibitemOpen
  \bibfield  {author} {\bibinfo {author} {\bibfnamefont {G.~A.}\ \bibnamefont {\'Alvarez}}, \bibinfo {author} {\bibfnamefont {A.~M.}\ \bibnamefont {Souza}},\ and\ \bibinfo {author} {\bibfnamefont {D.}~\bibnamefont {Suter}},\ }\bibfield  {title} {\bibinfo {title} {Iterative rotation scheme for robust dynamical decoupling},\ }\href {https://doi.org/10.1103/PhysRevA.85.052324} {\bibfield  {journal} {\bibinfo  {journal} {Phys. Rev. A}\ }\textbf {\bibinfo {volume} {85}},\ \bibinfo {pages} {052324} (\bibinfo {year} {2012})}\BibitemShut {NoStop}%
\bibitem [{\citenamefont {Viola}\ \emph {et~al.}(1999)\citenamefont {Viola}, \citenamefont {Knill},\ and\ \citenamefont {Lloyd}}]{viola1999dynamical}%
  \BibitemOpen
  \bibfield  {author} {\bibinfo {author} {\bibfnamefont {L.}~\bibnamefont {Viola}}, \bibinfo {author} {\bibfnamefont {E.}~\bibnamefont {Knill}},\ and\ \bibinfo {author} {\bibfnamefont {S.}~\bibnamefont {Lloyd}},\ }\bibfield  {title} {\bibinfo {title} {Dynamical decoupling of open quantum systems},\ }\href {https://doi.org/10.1103/PhysRevLett.82.2417} {\bibfield  {journal} {\bibinfo  {journal} {Phys. Rev. Lett.}\ }\textbf {\bibinfo {volume} {82}},\ \bibinfo {pages} {2417} (\bibinfo {year} {1999})}\BibitemShut {NoStop}%
\bibitem [{\citenamefont {Souza}\ \emph {et~al.}(2012{\natexlab{b}})\citenamefont {Souza}, \citenamefont {\'Alvarez},\ and\ \citenamefont {Suter}}]{souza2012effects}%
  \BibitemOpen
  \bibfield  {author} {\bibinfo {author} {\bibfnamefont {A.~M.}\ \bibnamefont {Souza}}, \bibinfo {author} {\bibfnamefont {G.~A.}\ \bibnamefont {\'Alvarez}},\ and\ \bibinfo {author} {\bibfnamefont {D.}~\bibnamefont {Suter}},\ }\bibfield  {title} {\bibinfo {title} {Effects of time-reversal symmetry in dynamical decoupling},\ }\href {https://doi.org/10.1103/PhysRevA.85.032306} {\bibfield  {journal} {\bibinfo  {journal} {Phys. Rev. A}\ }\textbf {\bibinfo {volume} {85}},\ \bibinfo {pages} {032306} (\bibinfo {year} {2012}{\natexlab{b}})}\BibitemShut {NoStop}%
\bibitem [{\citenamefont {Souza}\ \emph {et~al.}(2011)\citenamefont {Souza}, \citenamefont {\'Alvarez},\ and\ \citenamefont {Suter}}]{souza2011robust}%
  \BibitemOpen
  \bibfield  {author} {\bibinfo {author} {\bibfnamefont {A.~M.}\ \bibnamefont {Souza}}, \bibinfo {author} {\bibfnamefont {G.~A.}\ \bibnamefont {\'Alvarez}},\ and\ \bibinfo {author} {\bibfnamefont {D.}~\bibnamefont {Suter}},\ }\bibfield  {title} {\bibinfo {title} {Robust dynamical decoupling for quantum computing and quantum memory},\ }\href {https://doi.org/10.1103/PhysRevLett.106.240501} {\bibfield  {journal} {\bibinfo  {journal} {Phys. Rev. Lett.}\ }\textbf {\bibinfo {volume} {106}},\ \bibinfo {pages} {240501} (\bibinfo {year} {2011})}\BibitemShut {NoStop}%
\bibitem [{\citenamefont {Uhrig}(2007)}]{uhrig2007keeping}%
  \BibitemOpen
  \bibfield  {author} {\bibinfo {author} {\bibfnamefont {G.~S.}\ \bibnamefont {Uhrig}},\ }\bibfield  {title} {\bibinfo {title} {Keeping a quantum bit alive by optimized $\ensuremath{\pi}$-pulse sequences},\ }\href {https://doi.org/10.1103/PhysRevLett.98.100504} {\bibfield  {journal} {\bibinfo  {journal} {Phys. Rev. Lett.}\ }\textbf {\bibinfo {volume} {98}},\ \bibinfo {pages} {100504} (\bibinfo {year} {2007})}\BibitemShut {NoStop}%
\bibitem [{\citenamefont {Ezzell}\ \emph {et~al.}(2023)\citenamefont {Ezzell}, \citenamefont {Pokharel}, \citenamefont {Tewala}, \citenamefont {Quiroz},\ and\ \citenamefont {Lidar}}]{ezzell2023dynamical}%
  \BibitemOpen
  \bibfield  {author} {\bibinfo {author} {\bibfnamefont {N.}~\bibnamefont {Ezzell}}, \bibinfo {author} {\bibfnamefont {B.}~\bibnamefont {Pokharel}}, \bibinfo {author} {\bibfnamefont {L.}~\bibnamefont {Tewala}}, \bibinfo {author} {\bibfnamefont {G.}~\bibnamefont {Quiroz}},\ and\ \bibinfo {author} {\bibfnamefont {D.~A.}\ \bibnamefont {Lidar}},\ }\bibfield  {title} {\bibinfo {title} {Dynamical decoupling for superconducting qubits: A performance survey},\ }\href {https://doi.org/10.1103/PhysRevApplied.20.064027} {\bibfield  {journal} {\bibinfo  {journal} {Phys. Rev. Appl.}\ }\textbf {\bibinfo {volume} {20}},\ \bibinfo {pages} {064027} (\bibinfo {year} {2023})}\BibitemShut {NoStop}%
\bibitem [{\citenamefont {Farhi}\ \emph {et~al.}(2014)\citenamefont {Farhi}, \citenamefont {Goldstone},\ and\ \citenamefont {Gutmann}}]{farhi2014quantum}%
  \BibitemOpen
  \bibfield  {author} {\bibinfo {author} {\bibfnamefont {E.}~\bibnamefont {Farhi}}, \bibinfo {author} {\bibfnamefont {J.}~\bibnamefont {Goldstone}},\ and\ \bibinfo {author} {\bibfnamefont {S.}~\bibnamefont {Gutmann}},\ }\bibfield  {title} {\bibinfo {title} {A quantum approximate optimization algorithm},\ }\href@noop {} {\bibfield  {journal} {\bibinfo  {journal} {arXiv preprint arXiv:1411.4028}\ } (\bibinfo {year} {2014})}\BibitemShut {NoStop}%
\bibitem [{\citenamefont {Blekos}\ \emph {et~al.}(2024)\citenamefont {Blekos}, \citenamefont {Brand}, \citenamefont {Ceschini}, \citenamefont {Chou}, \citenamefont {Li}, \citenamefont {Pandya},\ and\ \citenamefont {Summer}}]{blekos2024review}%
  \BibitemOpen
  \bibfield  {author} {\bibinfo {author} {\bibfnamefont {K.}~\bibnamefont {Blekos}}, \bibinfo {author} {\bibfnamefont {D.}~\bibnamefont {Brand}}, \bibinfo {author} {\bibfnamefont {A.}~\bibnamefont {Ceschini}}, \bibinfo {author} {\bibfnamefont {C.-H.}\ \bibnamefont {Chou}}, \bibinfo {author} {\bibfnamefont {R.-H.}\ \bibnamefont {Li}}, \bibinfo {author} {\bibfnamefont {K.}~\bibnamefont {Pandya}},\ and\ \bibinfo {author} {\bibfnamefont {A.}~\bibnamefont {Summer}},\ }\bibfield  {title} {\bibinfo {title} {A review on quantum approximate optimization algorithm and its variants},\ }\href@noop {} {\bibfield  {journal} {\bibinfo  {journal} {Physics Reports}\ }\textbf {\bibinfo {volume} {1068}},\ \bibinfo {pages} {1} (\bibinfo {year} {2024})}\BibitemShut {NoStop}%
\bibitem [{\citenamefont {Misra-Spieldenner}\ \emph {et~al.}(2023)\citenamefont {Misra-Spieldenner}, \citenamefont {Bode}, \citenamefont {Schuhmacher}, \citenamefont {Stollenwerk}, \citenamefont {Bagrets},\ and\ \citenamefont {Wilhelm}}]{spieldenner2023mean}%
  \BibitemOpen
  \bibfield  {author} {\bibinfo {author} {\bibfnamefont {A.}~\bibnamefont {Misra-Spieldenner}}, \bibinfo {author} {\bibfnamefont {T.}~\bibnamefont {Bode}}, \bibinfo {author} {\bibfnamefont {P.~K.}\ \bibnamefont {Schuhmacher}}, \bibinfo {author} {\bibfnamefont {T.}~\bibnamefont {Stollenwerk}}, \bibinfo {author} {\bibfnamefont {D.}~\bibnamefont {Bagrets}},\ and\ \bibinfo {author} {\bibfnamefont {F.~K.}\ \bibnamefont {Wilhelm}},\ }\bibfield  {title} {\bibinfo {title} {Mean-field approximate optimization algorithm},\ }\href {https://doi.org/10.1103/PRXQuantum.4.030335} {\bibfield  {journal} {\bibinfo  {journal} {PRX Quantum}\ }\textbf {\bibinfo {volume} {4}},\ \bibinfo {pages} {030335} (\bibinfo {year} {2023})}\BibitemShut {NoStop}%
\bibitem [{\citenamefont {Zhou}\ \emph {et~al.}(2020)\citenamefont {Zhou}, \citenamefont {Wang}, \citenamefont {Choi}, \citenamefont {Pichler},\ and\ \citenamefont {Lukin}}]{zhou2020quantum}%
  \BibitemOpen
  \bibfield  {author} {\bibinfo {author} {\bibfnamefont {L.}~\bibnamefont {Zhou}}, \bibinfo {author} {\bibfnamefont {S.-T.}\ \bibnamefont {Wang}}, \bibinfo {author} {\bibfnamefont {S.}~\bibnamefont {Choi}}, \bibinfo {author} {\bibfnamefont {H.}~\bibnamefont {Pichler}},\ and\ \bibinfo {author} {\bibfnamefont {M.~D.}\ \bibnamefont {Lukin}},\ }\bibfield  {title} {\bibinfo {title} {Quantum approximate optimization algorithm: Performance, mechanism, and implementation on near-term devices},\ }\href@noop {} {\bibfield  {journal} {\bibinfo  {journal} {Physical Review X}\ }\textbf {\bibinfo {volume} {10}},\ \bibinfo {pages} {021067} (\bibinfo {year} {2020})}\BibitemShut {NoStop}%
\bibitem [{\citenamefont {Harrigan}\ \emph {et~al.}(2021)\citenamefont {Harrigan}, \citenamefont {Sung}, \citenamefont {Neeley}, \citenamefont {Satzinger}, \citenamefont {Arute}, \citenamefont {Arya}, \citenamefont {Atalaya}, \citenamefont {Bardin}, \citenamefont {Barends}, \citenamefont {Boixo} \emph {et~al.}}]{harrigan2021quantum}%
  \BibitemOpen
  \bibfield  {author} {\bibinfo {author} {\bibfnamefont {M.~P.}\ \bibnamefont {Harrigan}}, \bibinfo {author} {\bibfnamefont {K.~J.}\ \bibnamefont {Sung}}, \bibinfo {author} {\bibfnamefont {M.}~\bibnamefont {Neeley}}, \bibinfo {author} {\bibfnamefont {K.~J.}\ \bibnamefont {Satzinger}}, \bibinfo {author} {\bibfnamefont {F.}~\bibnamefont {Arute}}, \bibinfo {author} {\bibfnamefont {K.}~\bibnamefont {Arya}}, \bibinfo {author} {\bibfnamefont {J.}~\bibnamefont {Atalaya}}, \bibinfo {author} {\bibfnamefont {J.~C.}\ \bibnamefont {Bardin}}, \bibinfo {author} {\bibfnamefont {R.}~\bibnamefont {Barends}}, \bibinfo {author} {\bibfnamefont {S.}~\bibnamefont {Boixo}}, \emph {et~al.},\ }\bibfield  {title} {\bibinfo {title} {Quantum approximate optimization of non-planar graph problems on a planar superconducting processor},\ }\href@noop {} {\bibfield  {journal} {\bibinfo  {journal} {Nature Physics}\ }\textbf {\bibinfo {volume} {17}},\ \bibinfo {pages} {332} (\bibinfo {year} {2021})}\BibitemShut {NoStop}%
\bibitem [{\citenamefont {Niu}\ and\ \citenamefont {Todri-Sanial}(2022)}]{niu2022effects}%
  \BibitemOpen
  \bibfield  {author} {\bibinfo {author} {\bibfnamefont {S.}~\bibnamefont {Niu}}\ and\ \bibinfo {author} {\bibfnamefont {A.}~\bibnamefont {Todri-Sanial}},\ }\bibfield  {title} {\bibinfo {title} {Effects of dynamical decoupling and pulse-level optimizations on ibm quantum computers},\ }\href@noop {} {\bibfield  {journal} {\bibinfo  {journal} {IEEE Transactions on Quantum Engineering}\ }\textbf {\bibinfo {volume} {3}},\ \bibinfo {pages} {1} (\bibinfo {year} {2022})}\BibitemShut {NoStop}%
\bibitem [{\citenamefont {Das}\ \emph {et~al.}(2021)\citenamefont {Das}, \citenamefont {Tannu}, \citenamefont {Dangwal},\ and\ \citenamefont {Qureshi}}]{das2021adapt}%
  \BibitemOpen
  \bibfield  {author} {\bibinfo {author} {\bibfnamefont {P.}~\bibnamefont {Das}}, \bibinfo {author} {\bibfnamefont {S.}~\bibnamefont {Tannu}}, \bibinfo {author} {\bibfnamefont {S.}~\bibnamefont {Dangwal}},\ and\ \bibinfo {author} {\bibfnamefont {M.}~\bibnamefont {Qureshi}},\ }\bibfield  {title} {\bibinfo {title} {Adapt: Mitigating idling errors in qubits via adaptive dynamical decoupling},\ }in\ \href@noop {} {\emph {\bibinfo {booktitle} {MICRO-54: 54th Annual IEEE/ACM International Symposium on Microarchitecture}}}\ (\bibinfo {year} {2021})\ pp.\ \bibinfo {pages} {950--962}\BibitemShut {NoStop}%
\bibitem [{\citenamefont {Tong}\ \emph {et~al.}(2024)\citenamefont {Tong}, \citenamefont {Zhang},\ and\ \citenamefont {Pokharel}}]{tong2024empirical}%
  \BibitemOpen
  \bibfield  {author} {\bibinfo {author} {\bibfnamefont {C.}~\bibnamefont {Tong}}, \bibinfo {author} {\bibfnamefont {H.}~\bibnamefont {Zhang}},\ and\ \bibinfo {author} {\bibfnamefont {B.}~\bibnamefont {Pokharel}},\ }\bibfield  {title} {\bibinfo {title} {Empirical learning of dynamical decoupling on quantum processors},\ }\href@noop {} {\bibfield  {journal} {\bibinfo  {journal} {arXiv preprint arXiv:2403.02294}\ } (\bibinfo {year} {2024})}\BibitemShut {NoStop}%
\bibitem [{\citenamefont {Ji}\ \emph {et~al.}(2022)\citenamefont {Ji}, \citenamefont {Brandhofer},\ and\ \citenamefont {Polian}}]{ji2022calibration}%
  \BibitemOpen
  \bibfield  {author} {\bibinfo {author} {\bibfnamefont {Y.}~\bibnamefont {Ji}}, \bibinfo {author} {\bibfnamefont {S.}~\bibnamefont {Brandhofer}},\ and\ \bibinfo {author} {\bibfnamefont {I.}~\bibnamefont {Polian}},\ }\bibfield  {title} {\bibinfo {title} {Calibration-aware transpilation for variational quantum optimization},\ }in\ \href@noop {} {\emph {\bibinfo {booktitle} {2022 IEEE International Conference on Quantum Computing and Engineering (QCE)}}}\ (\bibinfo {organization} {IEEE},\ \bibinfo {year} {2022})\ pp.\ \bibinfo {pages} {204--214}\BibitemShut {NoStop}%
\bibitem [{\citenamefont {Gokhale}\ \emph {et~al.}(2020)\citenamefont {Gokhale}, \citenamefont {Javadi-Abhari}, \citenamefont {Earnest}, \citenamefont {Shi},\ and\ \citenamefont {Chong}}]{gokhale2020optimized}%
  \BibitemOpen
  \bibfield  {author} {\bibinfo {author} {\bibfnamefont {P.}~\bibnamefont {Gokhale}}, \bibinfo {author} {\bibfnamefont {A.}~\bibnamefont {Javadi-Abhari}}, \bibinfo {author} {\bibfnamefont {N.}~\bibnamefont {Earnest}}, \bibinfo {author} {\bibfnamefont {Y.}~\bibnamefont {Shi}},\ and\ \bibinfo {author} {\bibfnamefont {F.~T.}\ \bibnamefont {Chong}},\ }\bibfield  {title} {\bibinfo {title} {Optimized quantum compilation for near-term algorithms with openpulse},\ }in\ \href@noop {} {\emph {\bibinfo {booktitle} {2020 53rd Annual IEEE/ACM International Symposium on Microarchitecture (MICRO)}}}\ (\bibinfo {organization} {IEEE},\ \bibinfo {year} {2020})\ pp.\ \bibinfo {pages} {186--200}\BibitemShut {NoStop}%
\bibitem [{\citenamefont {Leymann}\ and\ \citenamefont {Barzen}(2020)}]{leymann2020bitter}%
  \BibitemOpen
  \bibfield  {author} {\bibinfo {author} {\bibfnamefont {F.}~\bibnamefont {Leymann}}\ and\ \bibinfo {author} {\bibfnamefont {J.}~\bibnamefont {Barzen}},\ }\bibfield  {title} {\bibinfo {title} {The bitter truth about gate-based quantum algorithms in the nisq era},\ }\href@noop {} {\bibfield  {journal} {\bibinfo  {journal} {Quantum Science and Technology}\ }\textbf {\bibinfo {volume} {5}},\ \bibinfo {pages} {044007} (\bibinfo {year} {2020})}\BibitemShut {NoStop}%
\bibitem [{\citenamefont {Huang}\ \emph {et~al.}(2023)\citenamefont {Huang}, \citenamefont {Xu}, \citenamefont {Guo}, \citenamefont {Tian}, \citenamefont {Wei}, \citenamefont {Sun}, \citenamefont {Bao},\ and\ \citenamefont {Long}}]{huang2023near}%
  \BibitemOpen
  \bibfield  {author} {\bibinfo {author} {\bibfnamefont {H.-L.}\ \bibnamefont {Huang}}, \bibinfo {author} {\bibfnamefont {X.-Y.}\ \bibnamefont {Xu}}, \bibinfo {author} {\bibfnamefont {C.}~\bibnamefont {Guo}}, \bibinfo {author} {\bibfnamefont {G.}~\bibnamefont {Tian}}, \bibinfo {author} {\bibfnamefont {S.-J.}\ \bibnamefont {Wei}}, \bibinfo {author} {\bibfnamefont {X.}~\bibnamefont {Sun}}, \bibinfo {author} {\bibfnamefont {W.-S.}\ \bibnamefont {Bao}},\ and\ \bibinfo {author} {\bibfnamefont {G.-L.}\ \bibnamefont {Long}},\ }\bibfield  {title} {\bibinfo {title} {Near-term quantum computing techniques: Variational quantum algorithms, error mitigation, circuit compilation, benchmarking and classical simulation},\ }\href@noop {} {\bibfield  {journal} {\bibinfo  {journal} {Science China Physics, Mechanics \& Astronomy}\ }\textbf {\bibinfo {volume} {66}},\ \bibinfo {pages} {250302} (\bibinfo {year} {2023})}\BibitemShut {NoStop}%
\bibitem [{\citenamefont {Ji}\ \emph {et~al.}(2023{\natexlab{a}})\citenamefont {Ji}, \citenamefont {Chen}, \citenamefont {Polian},\ and\ \citenamefont {Ban}}]{ji2023algorithm}%
  \BibitemOpen
  \bibfield  {author} {\bibinfo {author} {\bibfnamefont {Y.}~\bibnamefont {Ji}}, \bibinfo {author} {\bibfnamefont {X.}~\bibnamefont {Chen}}, \bibinfo {author} {\bibfnamefont {I.}~\bibnamefont {Polian}},\ and\ \bibinfo {author} {\bibfnamefont {Y.}~\bibnamefont {Ban}},\ }\bibfield  {title} {\bibinfo {title} {Algorithm-oriented qubit mapping for variational quantum algorithms},\ }\href@noop {} {\bibfield  {journal} {\bibinfo  {journal} {arXiv preprint arXiv:2310.09826}\ } (\bibinfo {year} {2023}{\natexlab{a}})}\BibitemShut {NoStop}%
\bibitem [{\citenamefont {Ji}\ \emph {et~al.}(2023{\natexlab{b}})\citenamefont {Ji}, \citenamefont {Koenig},\ and\ \citenamefont {Polian}}]{ji2023improving}%
  \BibitemOpen
  \bibfield  {author} {\bibinfo {author} {\bibfnamefont {Y.}~\bibnamefont {Ji}}, \bibinfo {author} {\bibfnamefont {K.~F.}\ \bibnamefont {Koenig}},\ and\ \bibinfo {author} {\bibfnamefont {I.}~\bibnamefont {Polian}},\ }\bibfield  {title} {\bibinfo {title} {Improving the performance of digitized counterdiabatic quantum optimization via algorithm-oriented qubit mapping},\ }\href@noop {} {\bibfield  {journal} {\bibinfo  {journal} {arXiv preprint arXiv:2311.14624}\ } (\bibinfo {year} {2023}{\natexlab{b}})}\BibitemShut {NoStop}%
\bibitem [{\citenamefont {Ji}\ \emph {et~al.}(2023{\natexlab{c}})\citenamefont {Ji}, \citenamefont {Koenig},\ and\ \citenamefont {Polian}}]{ji2023optimizing}%
  \BibitemOpen
  \bibfield  {author} {\bibinfo {author} {\bibfnamefont {Y.}~\bibnamefont {Ji}}, \bibinfo {author} {\bibfnamefont {K.~F.}\ \bibnamefont {Koenig}},\ and\ \bibinfo {author} {\bibfnamefont {I.}~\bibnamefont {Polian}},\ }\bibfield  {title} {\bibinfo {title} {Optimizing quantum algorithms on bipotent architectures},\ }\href {https://doi.org/10.1103/PhysRevA.108.022610} {\bibfield  {journal} {\bibinfo  {journal} {Phys. Rev. A}\ }\textbf {\bibinfo {volume} {108}},\ \bibinfo {pages} {022610} (\bibinfo {year} {2023}{\natexlab{c}})}\BibitemShut {NoStop}%
\bibitem [{\citenamefont {Vartiainen}\ \emph {et~al.}(2004)\citenamefont {Vartiainen}, \citenamefont {M\"ott\"onen},\ and\ \citenamefont {Salomaa}}]{vartiainen2004efficient}%
  \BibitemOpen
  \bibfield  {author} {\bibinfo {author} {\bibfnamefont {J.~J.}\ \bibnamefont {Vartiainen}}, \bibinfo {author} {\bibfnamefont {M.}~\bibnamefont {M\"ott\"onen}},\ and\ \bibinfo {author} {\bibfnamefont {M.~M.}\ \bibnamefont {Salomaa}},\ }\bibfield  {title} {\bibinfo {title} {Efficient decomposition of quantum gates},\ }\href {https://doi.org/10.1103/PhysRevLett.92.177902} {\bibfield  {journal} {\bibinfo  {journal} {Phys. Rev. Lett.}\ }\textbf {\bibinfo {volume} {92}},\ \bibinfo {pages} {177902} (\bibinfo {year} {2004})}\BibitemShut {NoStop}%
\bibitem [{\citenamefont {Baker}\ and\ \citenamefont {Radha}(2022)}]{baker2022wasserstein}%
  \BibitemOpen
  \bibfield  {author} {\bibinfo {author} {\bibfnamefont {J.~S.}\ \bibnamefont {Baker}}\ and\ \bibinfo {author} {\bibfnamefont {S.~K.}\ \bibnamefont {Radha}},\ }\href {https://doi.org/10.48550/arXiv.2202.06782} {\bibinfo {title} {Wasserstein solution quality and the quantum approximate optimization algorithm: A portfolio optimization case study}} (\bibinfo {year} {2022}),\ \Eprint {https://arxiv.org/abs/2202.06782} {arXiv:2202.06782} \BibitemShut {NoStop}%
\bibitem [{\citenamefont {Brandhofer}\ \emph {et~al.}(2022)\citenamefont {Brandhofer}, \citenamefont {Braun}, \citenamefont {Dehn}, \citenamefont {Hellstern}, \citenamefont {H{\"u}ls}, \citenamefont {Ji}, \citenamefont {Polian}, \citenamefont {Bhatia},\ and\ \citenamefont {Wellens}}]{brandhofer2022benchmarking}%
  \BibitemOpen
  \bibfield  {author} {\bibinfo {author} {\bibfnamefont {S.}~\bibnamefont {Brandhofer}}, \bibinfo {author} {\bibfnamefont {D.}~\bibnamefont {Braun}}, \bibinfo {author} {\bibfnamefont {V.}~\bibnamefont {Dehn}}, \bibinfo {author} {\bibfnamefont {G.}~\bibnamefont {Hellstern}}, \bibinfo {author} {\bibfnamefont {M.}~\bibnamefont {H{\"u}ls}}, \bibinfo {author} {\bibfnamefont {Y.}~\bibnamefont {Ji}}, \bibinfo {author} {\bibfnamefont {I.}~\bibnamefont {Polian}}, \bibinfo {author} {\bibfnamefont {A.~S.}\ \bibnamefont {Bhatia}},\ and\ \bibinfo {author} {\bibfnamefont {T.}~\bibnamefont {Wellens}},\ }\bibfield  {title} {\bibinfo {title} {Benchmarking the performance of portfolio optimization with qaoa},\ }\href@noop {} {\bibfield  {journal} {\bibinfo  {journal} {Quantum Information Processing}\ }\textbf {\bibinfo {volume} {22}},\ \bibinfo {pages} {25} (\bibinfo {year} {2022})}\BibitemShut {NoStop}%
\bibitem [{\citenamefont {Egger}\ \emph {et~al.}(2020)\citenamefont {Egger}, \citenamefont {Gambella}, \citenamefont {Marecek}, \citenamefont {McFaddin}, \citenamefont {Mevissen}, \citenamefont {Raymond}, \citenamefont {Simonetto}, \citenamefont {Woerner},\ and\ \citenamefont {Yndurain}}]{egger2020quantum}%
  \BibitemOpen
  \bibfield  {author} {\bibinfo {author} {\bibfnamefont {D.~J.}\ \bibnamefont {Egger}}, \bibinfo {author} {\bibfnamefont {C.}~\bibnamefont {Gambella}}, \bibinfo {author} {\bibfnamefont {J.}~\bibnamefont {Marecek}}, \bibinfo {author} {\bibfnamefont {S.}~\bibnamefont {McFaddin}}, \bibinfo {author} {\bibfnamefont {M.}~\bibnamefont {Mevissen}}, \bibinfo {author} {\bibfnamefont {R.}~\bibnamefont {Raymond}}, \bibinfo {author} {\bibfnamefont {A.}~\bibnamefont {Simonetto}}, \bibinfo {author} {\bibfnamefont {S.}~\bibnamefont {Woerner}},\ and\ \bibinfo {author} {\bibfnamefont {E.}~\bibnamefont {Yndurain}},\ }\bibfield  {title} {\bibinfo {title} {Quantum computing for finance: State-of-the-art and future prospects},\ }\href {https://doi.org/10.1109/TQE.2020.3030314} {\bibfield  {journal} {\bibinfo  {journal} {IEEE Transactions on Quantum Engineering}\ }\textbf {\bibinfo {volume} {1}},\ \bibinfo {pages} {1} (\bibinfo {year} {2020})}\BibitemShut {NoStop}%
\bibitem [{\citenamefont {Cerezo}\ \emph {et~al.}(2021)\citenamefont {Cerezo}, \citenamefont {Arrasmith}, \citenamefont {Babbush}, \citenamefont {Benjamin}, \citenamefont {Endo}, \citenamefont {Fujii}, \citenamefont {McClean}, \citenamefont {Mitarai}, \citenamefont {Yuan}, \citenamefont {Cincio} \emph {et~al.}}]{cerezo2021variational}%
  \BibitemOpen
  \bibfield  {author} {\bibinfo {author} {\bibfnamefont {M.}~\bibnamefont {Cerezo}}, \bibinfo {author} {\bibfnamefont {A.}~\bibnamefont {Arrasmith}}, \bibinfo {author} {\bibfnamefont {R.}~\bibnamefont {Babbush}}, \bibinfo {author} {\bibfnamefont {S.~C.}\ \bibnamefont {Benjamin}}, \bibinfo {author} {\bibfnamefont {S.}~\bibnamefont {Endo}}, \bibinfo {author} {\bibfnamefont {K.}~\bibnamefont {Fujii}}, \bibinfo {author} {\bibfnamefont {J.~R.}\ \bibnamefont {McClean}}, \bibinfo {author} {\bibfnamefont {K.}~\bibnamefont {Mitarai}}, \bibinfo {author} {\bibfnamefont {X.}~\bibnamefont {Yuan}}, \bibinfo {author} {\bibfnamefont {L.}~\bibnamefont {Cincio}}, \emph {et~al.},\ }\bibfield  {title} {\bibinfo {title} {Variational quantum algorithms},\ }\href@noop {} {\bibfield  {journal} {\bibinfo  {journal} {Nature Reviews Physics}\ }\textbf {\bibinfo {volume} {3}},\ \bibinfo {pages} {625} (\bibinfo {year} {2021})}\BibitemShut {NoStop}%
\bibitem [{\citenamefont {Contributors}(2023)}]{qiskit}%
  \BibitemOpen
  \bibfield  {author} {\bibinfo {author} {\bibfnamefont {Q.}~\bibnamefont {Contributors}},\ }\bibfield  {title} {\bibinfo {title} {Qiskit: An open-source framework for quantum computing},\ }\href@noop {} {\bibfield  {journal} {\bibinfo  {journal} {Zenodo: Geneva, Switzerland}\ } (\bibinfo {year} {2023})}\BibitemShut {NoStop}%
\bibitem [{\citenamefont {Sivarajah}\ \emph {et~al.}(2020)\citenamefont {Sivarajah}, \citenamefont {Dilkes}, \citenamefont {Cowtan}, \citenamefont {Simmons}, \citenamefont {Edgington},\ and\ \citenamefont {Duncan}}]{sivarajah2020tket}%
  \BibitemOpen
  \bibfield  {author} {\bibinfo {author} {\bibfnamefont {S.}~\bibnamefont {Sivarajah}}, \bibinfo {author} {\bibfnamefont {S.}~\bibnamefont {Dilkes}}, \bibinfo {author} {\bibfnamefont {A.}~\bibnamefont {Cowtan}}, \bibinfo {author} {\bibfnamefont {W.}~\bibnamefont {Simmons}}, \bibinfo {author} {\bibfnamefont {A.}~\bibnamefont {Edgington}},\ and\ \bibinfo {author} {\bibfnamefont {R.}~\bibnamefont {Duncan}},\ }\bibfield  {title} {\bibinfo {title} {t|ket⟩: a retargetable compiler for nisq devices},\ }\href {https://doi.org/10.1088/2058-9565/ab8e92} {\bibfield  {journal} {\bibinfo  {journal} {Quantum Science and Technology}\ }\textbf {\bibinfo {volume} {6}},\ \bibinfo {pages} {014003} (\bibinfo {year} {2020})}\BibitemShut {NoStop}%
\bibitem [{\citenamefont {Peruzzo}\ \emph {et~al.}(2014)\citenamefont {Peruzzo}, \citenamefont {McClean}, \citenamefont {Shadbolt}, \citenamefont {Yung}, \citenamefont {Zhou}, \citenamefont {Love}, \citenamefont {Aspuru-Guzik},\ and\ \citenamefont {O’brien}}]{peruzzo2014variational}%
  \BibitemOpen
  \bibfield  {author} {\bibinfo {author} {\bibfnamefont {A.}~\bibnamefont {Peruzzo}}, \bibinfo {author} {\bibfnamefont {J.}~\bibnamefont {McClean}}, \bibinfo {author} {\bibfnamefont {P.}~\bibnamefont {Shadbolt}}, \bibinfo {author} {\bibfnamefont {M.-H.}\ \bibnamefont {Yung}}, \bibinfo {author} {\bibfnamefont {X.-Q.}\ \bibnamefont {Zhou}}, \bibinfo {author} {\bibfnamefont {P.~J.}\ \bibnamefont {Love}}, \bibinfo {author} {\bibfnamefont {A.}~\bibnamefont {Aspuru-Guzik}},\ and\ \bibinfo {author} {\bibfnamefont {J.~L.}\ \bibnamefont {O’brien}},\ }\bibfield  {title} {\bibinfo {title} {A variational eigenvalue solver on a photonic quantum processor},\ }\href@noop {} {\bibfield  {journal} {\bibinfo  {journal} {Nature communications}\ }\textbf {\bibinfo {volume} {5}},\ \bibinfo {pages} {1} (\bibinfo {year} {2014})}\BibitemShut {NoStop}%
\bibitem [{\citenamefont {Tilly}\ \emph {et~al.}(2022)\citenamefont {Tilly}, \citenamefont {Chen}, \citenamefont {Cao}, \citenamefont {Picozzi}, \citenamefont {Setia}, \citenamefont {Li}, \citenamefont {Grant}, \citenamefont {Wossnig}, \citenamefont {Rungger}, \citenamefont {Booth} \emph {et~al.}}]{tilly2022variational}%
  \BibitemOpen
  \bibfield  {author} {\bibinfo {author} {\bibfnamefont {J.}~\bibnamefont {Tilly}}, \bibinfo {author} {\bibfnamefont {H.}~\bibnamefont {Chen}}, \bibinfo {author} {\bibfnamefont {S.}~\bibnamefont {Cao}}, \bibinfo {author} {\bibfnamefont {D.}~\bibnamefont {Picozzi}}, \bibinfo {author} {\bibfnamefont {K.}~\bibnamefont {Setia}}, \bibinfo {author} {\bibfnamefont {Y.}~\bibnamefont {Li}}, \bibinfo {author} {\bibfnamefont {E.}~\bibnamefont {Grant}}, \bibinfo {author} {\bibfnamefont {L.}~\bibnamefont {Wossnig}}, \bibinfo {author} {\bibfnamefont {I.}~\bibnamefont {Rungger}}, \bibinfo {author} {\bibfnamefont {G.~H.}\ \bibnamefont {Booth}}, \emph {et~al.},\ }\bibfield  {title} {\bibinfo {title} {The variational quantum eigensolver: a review of methods and best practices},\ }\href {https://doi.org/https://doi.org/10.1016/j.physrep.2022.08.003} {\bibfield  {journal} {\bibinfo  {journal} {Physics Reports}\ }\textbf {\bibinfo {volume} {986}},\ \bibinfo {pages} {1} (\bibinfo {year} {2022})}\BibitemShut {NoStop}%
\bibitem [{\citenamefont {Li}\ and\ \citenamefont {Benjamin}(2017)}]{li2017efficient}%
  \BibitemOpen
  \bibfield  {author} {\bibinfo {author} {\bibfnamefont {Y.}~\bibnamefont {Li}}\ and\ \bibinfo {author} {\bibfnamefont {S.~C.}\ \bibnamefont {Benjamin}},\ }\bibfield  {title} {\bibinfo {title} {Efficient variational quantum simulator incorporating active error minimization},\ }\href {https://doi.org/10.1103/PhysRevX.7.021050} {\bibfield  {journal} {\bibinfo  {journal} {Phys. Rev. X}\ }\textbf {\bibinfo {volume} {7}},\ \bibinfo {pages} {021050} (\bibinfo {year} {2017})}\BibitemShut {NoStop}%
\bibitem [{\citenamefont {Temme}\ \emph {et~al.}(2017)\citenamefont {Temme}, \citenamefont {Bravyi},\ and\ \citenamefont {Gambetta}}]{temme2017error}%
  \BibitemOpen
  \bibfield  {author} {\bibinfo {author} {\bibfnamefont {K.}~\bibnamefont {Temme}}, \bibinfo {author} {\bibfnamefont {S.}~\bibnamefont {Bravyi}},\ and\ \bibinfo {author} {\bibfnamefont {J.~M.}\ \bibnamefont {Gambetta}},\ }\bibfield  {title} {\bibinfo {title} {Error mitigation for short-depth quantum circuits},\ }\href {https://doi.org/10.1103/PhysRevLett.119.180509} {\bibfield  {journal} {\bibinfo  {journal} {Phys. Rev. Lett.}\ }\textbf {\bibinfo {volume} {119}},\ \bibinfo {pages} {180509} (\bibinfo {year} {2017})}\BibitemShut {NoStop}%
\end{thebibliography}%

\end{document}